\documentclass[letterpaper,twocolumn,10pt]{article}
\usepackage{usenix2019_v3,epsfig,endnotes,booktabs,url,graphicx,amsmath}
\usepackage{tikz,threeparttable,makecell,hyperref}
\usetikzlibrary{positioning}
\usetikzlibrary{arrows}

\usepackage{xcolor}
\usepackage{multirow}
\usepackage{tabularx}
\usepackage{textcomp}
\usepackage{microtype}
\usepackage{float}
\usepackage{verbatim} 

\def\ORG{{\texttt {Org}}}

\newenvironment{myquote}[1]%
  {\list{}{\leftmargin=#1\rightmargin=#1}\item[]}%
  {\endlist}

\begin{document}

\date{}

\title{\Large \bf Cognitive Triaging of Phishing Attacks}

\author{
{\rm Amber van der Heijden}\\
a.v.d.heijden@student.tue.nl\\
Eindhoven University of Technology
\and
{\rm Luca Allodi}\\
l.allodi@tue.nl\\
Eindhoven University of Technology
} 

\maketitle

\thispagestyle{empty}
\pagenumbering{gobble}

\subsection*{Abstract}
In this paper we employ quantitative measurements of \emph{cognitive vulnerability triggers} in phishing emails to predict the degree of success of an attack. To achieve this we rely on the cognitive psychology literature and develop an automated and fully quantitative method based on machine learning and econometrics to construct a triaging mechanism built around the cognitive features of a phishing email; we showcase our approach relying on data from the anti-phishing division of a large financial organization in Europe. Our evaluation shows empirically that an effective triaging mechanism for phishing success can be put in place by response teams to effectively prioritize remediation efforts (e.g. domain takedowns), by first acting on those attacks that are more likely to collect high response rates from potential victims.

\section{Introduction}
\label{sec:intro}
Phishing attacks represent a significant threat to organizations and their customers~\cite{PhishLabs2018}. {The problem of phishing detection has been addressed multiple times in the literature~\cite{ho2017detecting, Jain2017, Moghimi2016}, yet classification is only part of the issue. A timely and efficient \emph{reaction} to phishing attempts (e.g. performing takedown actions on phishing domains, blacklisting, or notifying customers) could save hundreds or thousands of customers from fraud or theft, and associated costs for all involved stakeholders.
For this reason, most `large enough' organizations operate a phishing-response team whose task is to promptly investigate potential impacts, identify rogue domains and attack vectors, and act to contain or neutralize the attack~\cite{cichonski2012computer}. The size of this effort often requires the full time operation of several experts within the response team~\cite{shah2019understanding}.}

Unfortunately, these teams currently lack of an objective and quantitative way of prioritizing response activities, which can lead to large inefficiencies in the response process. {Technical mechanisms are often in place to \emph{a-posteriori} quantify the success of a phishing attack, but these are technically limited to attacks `in scope' of the measuring mechanism (e.g. evaluating the requests for internal resources received by the organization's servers and originating from remote domains) and, importantly, cannot \emph{predict} how successful the attack is likely to be if no immediate mitigation is put in place. 

Key to {predicting} phishing success is the likelihood that a human will \emph{comply} with whatever instruction is in the phishing email.
} 
Cialdini pioneered the definition of `\emph{principles of influence}', namely \texttt{Reciprocity}, \texttt{Consistency}, \texttt{Social Proof}, \texttt{Authority}, \texttt{Liking}, and \texttt{Scarcity} as `cognitive triggers' that, once engaged, can greatly impact the likelihood of a human's decision to comply with what he or she is being requested to do~\cite{Cialdini2007}. {These principles have been used as a theoretical framework to investigate persuasion in different domains, such as sales and marketing~\cite{Cialdini2002}, organizational behaviour~\cite{Robertson2013}, and wellbeing~\cite{Tay2013}, as well as being linked to phishing effectiveness~\cite{Wash2018,Workman2008} in (synthetic) experimental settings~\cite{Wright2014}; however no means to automatically measure the cognitive features of a phishing email, and estimate their relation to phishing success `in the wild', currently exists.}  

In this paper we employ techniques from natural language processing and econometrics to build a method and estimation process to measure cognitive triggers in phishing emails, and to build a \emph{cognitive triaging model} of how successful an attack can be expected to be. We demonstrate empirically that the resulting estimations can be used to efficiently prioritize phishing response actions, by addressing first the (few) attacks that are likely to be highly successful.
To do this, we extensively analyze more than eighty thousand phishing emails received by the anti-phishing division of a very large European financial organization, quantify the `cognitive vulnerability triggers' embedded in the attacks, and relate them to the number of accesses to the remote phish domain that the anti-phishing division measured. This allows us to empirically derive a triaging model that, only based on cognitive features of the incoming phishing email, can predict how many `clicks' it can be expected to generate. 
\vspace{-0.18in}
\paragraph{Scope and contribution of this work.} With this work we aim at building a principled analysis that explains \emph{why} one can expect a certain phishing email to be successful, as opposed to building a {method} that `blindly' maps mail bodies to success of attack. {Importantly, with this work we do \emph{not} aim to build a classifier to distinguish phishing from non-phishing emails; instead, we propose a method to \emph{predict} to what extent a \emph{known} phishing attack can be expected to lure users in falling for it.} 
Our contributions can be summarized as follows:
\vspace{-0.08in}

\begin{itemize}
    \setlength{\itemsep}{1pt}
  \setlength{\parskip}{0pt}
  \setlength{\parsep}{0pt}
\item we provide the first empirical analysis of cognitive vulnerabilities
as exploited in the wild by attackers launching phishing attacks;
\item we employ a robust measurement methodology to identify cognitive vulnerability triggers in phishing emails, using supervised Latent Dirichlet Allocation, and a set of bootstrapped econometric simulations to build robust estimations
of model coefficients and predictions;
\item we show empirically the correlation between exploited cognitive factors and spoofed \texttt{From:} addresses with an objective evaluation of phishing success;
\item we quantitatively show that triaging phishing emails to prioritize remediation action is possible and effective in an operational setting.
\end{itemize}\vspace{-0.08in}


 This paper proceeds as follows: Section~\ref{sec:background} sets the background for this work in both the cognitive psychology and information security literature; Section~\ref{sec:maethodology} details the employed data and methodology, and Section~\ref{sec:expl} 
 reports the exploratory and cognitive analysis of the data. The cognitive model and predictions are presented in Section~\ref{sec:modelling}. Section~\ref{sec:discussion} provides a discussion of our results, and Section~\ref{sec:conclusions} concludes the paper.

\section{Background and Related Work}
\label{sec:background}

The general objective of a phishing attack is to convince a target to comply with a request, such as clicking a link to a phishing domain, downloading malware, or providing personal credentials. The effectiveness of these attacks significantly relies on how quickly the message can generate the desired response {\cite{Wright2014}}. 
Moreover, both cognitive~\cite{Wright2010,Workman2008} and technical \cite{Mao2013, Mishr2014,Rosiello2007} features are employed to lure users into falling for the phish and are known to be relevant to explain phishing effectiveness.

\subsection{Cognitive characterizations}
\label{sec:backgroundcog}

\textbf{Believability.} Phishers apply several techniques to increase believability of their phishing messages. For example, they may craft their phishing messages to resemble communications of the impersonated organizations as closely as possible \cite{Wright2010}. {This is commonly done by duplicating the look and feel of these communications by including logos and other branded graphics extracted from their legitimate counterparts, and by adopting a formal writing style \cite{Hale2015}.} Furthermore, the context of phishing messages is generally highly personalized to appeal to the targeted population~\cite{Wright2014}. These practices are enhanced by more technical measures, such as spoofing of the phishing source address, and the use of shortened URLs to hide the destination of the embedded phishing link~\cite{LePage2018}.

\textbf{Persuasiveness.} Persuasiveness is associated with the text content of the email. These techniques work by exploiting fundamental vulnerabilities of human cognition \cite{Mitnick2003} that can be explained by `shortcuts' in human cognitive processes that determine decisions on the basis of previous experiences, biases, or beliefs~\cite{Stanovich2000}. 
 Despite the clear benefits of these mental-shortcuts, they can result in irrational decision-making as well \cite{Tversky1975}. Cialdini \cite{Cialdini2007} identified several principles that explain how these mental shortcuts can be exploited for the persuasion of others {(e.g. for marketing purposes). Indeed, these principles are applied regularly in multiple domains, including marketing (e.g. to purchase a product or solution)~\cite{Cialdini2002}, organizational behaviour (e.g. to comply to policies)~\cite{Robertson2013}, and health and wellbeing (e.g. to adopt healthy lifestyles)~\cite{Tay2013}. As these are foundational to human decision-making processes~\cite{o2008elaboration}, these principles may not be effectively applied to \emph{distinguish} legitimate from illegitimate resources (e.g. a website, email, or conversation): any activity aiming at `influencing' one's behaviour (that being through spam or organization policies, phishing or advertisement) will employ some variation of these principles. On the other hand, these provide a solid foundation to evaluate how \emph{effective} an attempt at convincing a human can be expected to be.}
 Table \ref{table:phishing example} 
\begin{table*}[t]
\footnotesize
\centering
\caption{Definitions and examples of Cialdini's principles of influence in phishing emails}
\label{table:phishing example}
\begin{threeparttable}
\begin{tabular}{@{}p{1.5cm}p{9cm}p{6.5cm}@{}}
\toprule
Principle       & Definition \cite{Cialdini2007}                                                                                          & Phishing text example \tnote{2}                                                                                                                                                                                        \\ \midrule
\texttt{Reciprocity}                & Tendency to feel obliged to repay favours from others.``I do something for you, you do something for me."                                & ``While we work hard to keep our network secure, we're asking you to help us keep your account safe.''                              \\ \addlinespace[0.1cm]
\texttt{Consistency} & Tendency to behave in a way consistent with past decisions and behaviours. After committing to a certain view, company or product, people will act in accordance with those commitments.                   & ``You agreed to the terms and conditions before using our service, so we ask you to stop all activities that violate them. Click here to unflag your account for suspension.'' \\ \addlinespace[0.1cm]
\texttt{Social Proof}               & Tendency to reference the behaviour of others, by using the majority behaviour to guide their own actions.             & ``We are introducing new security features to our services. All customers must get their accounts verified again.''                                                                                             \\ \addlinespace[0.1cm]
\texttt{Authority}                  & Tendency to obey people in authoritative positions, following from the possibility of punishment for not complying with the authoritative requests.                                      &\makecell[tl]{``Best regards,\\  Executive Vice President of \textless{}company name\textgreater''}                                                                     \\ \addlinespace[0.1cm]
\texttt{Liking}                     & Preference for saying ``yes'' to the requests of people they know and like. People are programmed to like others who like them back and who are similar to them.        & ``We care for our customers and their online security. Confirm your identity .. so we can continue protecting you."                                                                             \\ \addlinespace[0.1cm]
\texttt{Scarcity}                   & Tendency to assign more value to items and opportunities when their availability is limited, not to waste the opportunity.& ``If your account information is not updated within 48 hours then your ability to access your account will be restricted."                                                                                     \\ \addlinespace[0.1cm] \bottomrule
\end{tabular}%
\begin{tablenotes}
  \footnotesize
  \item[2] Examples drawn from anti-phishing database at \url{http://www.millersmiles.co.uk}.
\end{tablenotes}
\end{threeparttable}
\end{table*}
provides examples and definitions of these principles.

Cialdini's principles of persuasion are strongly related to the successfulness of face-to-face social engineering efforts in the real world \cite{Sagarin2012} as well. 
Akbar \cite{Akbar2014} performed a quantitative analysis on 207 unique phishing emails to identify the application of Cialdini's persuasion principles in phishing emails. The results show the \texttt{Authority}, \texttt{Scarcity} and \texttt{Liking} principles to be most popular. A similar study was performed by Ferreira et al. \cite{Ferreira2015}, who found the \text{Liking} principle to be most popularly used, followed distantly by the principles of \texttt{Scarcity} and \texttt{Authority}. Differences can be explained by different experimental settings and application domains. 
Several other studies \cite{Wright2014, Butavicius2016, Ferreira20152} have addressed the prevalence and efficacy of Cialdini's principles in phishing attacks. Others have evaluated phishing campaigns against specific users~\cite{le2014look}, discussing some of the techniques used by phishers to lure their victims. Unlike these works, we integrate quantitative measures of cognitive attacks and measures of phishing success to predict attack effectiveness in operational settings. 

\subsection{{Phishing effectiveness}}
\label{sec:phisheff}
{

Previous work considered the inclusion of forged quality marks, images, and logos from trusted organizations as well as other signals of credibility as means to increase the effectiveness of a phishing attack~\cite{Dhamija2006}. Other more technical measures are employed to enhance the credibility of phishing as well, for example spoofing of the source email address, adoption of HTTPS instead of HTTP to convince the user the webpage is `safe'~\cite{PhishLabs2018}, or cloning of the original webpage. Several works have considered such visual similarities between phishing landing pages and their legitimate counterparts based on different features, including DOM tree structures~\cite{Rosiello2007}, CSS styling~\cite{Mao2013, Mishr2014}, content signatures \cite{Afroz2011, Huang2010}, and pixel and/or image properties~\cite{Chen2007, Dunlop2010}. Whereas these technical features constitute additional relevant information for the identification of a phishing attack, in this study we focus on the cognitive attacks embedded in an email text (as opposed to the visual clues included in a landing webpage) that affect the human decision making. Additionally, a number of user-studies has been conducted on the impact of client-side detection-assistance tools~\cite{Wu2006, kumaraguru2009school} and how people evaluate phishing web pages~\cite{Dhamija2006}. Various phishing detection mechanisms have been proposed based on technical features such as signatures of user email behaviour~\cite{stringhini2015ain}, email-header properties \cite{ho2017detecting}, impersonation limitations of attackers~\cite{marchal2017off}, search engine rankings~\cite{Liu2010}, and botnet effects~\cite{Pearce2014}. Additionally, \cite{marchal2018designing} presents a set of research guidelines for design and evaluation of such detection systems. These works have predominantly focused on the detection of phishing domains and emails by means of technical traces in order to prevent phishing attacks from happening in the first place. Unlike these studies, we focus on the evaluation of the potential of those attacks that, despite the countermeasures in place, make it through and must be timely addressed.

On the cognitive-side we can consider the impact of user demographics. Oliveira et al.~\cite{Oliveira2017} found age to be an important feature, finding younger adults to be more susceptible to  \texttt{Scarcity}, whereas older adults were more susceptible to \texttt{Reciprocity}. Other results of this study indicate the relevance of gender by finding older women to be most susceptible of all of the studied user groups. Furthermore, Wash and Cooper~\cite{Wash2018} demonstrated the impact of message presentation by showing how phishing training methods based on giving facts-and-advice were more effective when presented by an expert figure (\texttt{Authority}), whereas methods based on personal stories benefited more from presentation by people perceived as similar to the user (\texttt{Liking}). In the context of social media, user activity, consumption behaviour, and clicking norms in the social network were found to be important factors for phishing success \cite{Redmiles2018}. As opposed to focusing on the characteristics of the individuals that receive the phishing (as this information for the population of customers is generally unknown to organizations, or may be impossible to collect due to legal and ethical challenges), in this work we consider the expected aggregate responses of the phishing recipients as a function of the phishing emails.
}

\section{Methodology and Data collection}
\label{sec:maethodology}

\begin{figure}
\centering
\includegraphics[width=0.49\textwidth]{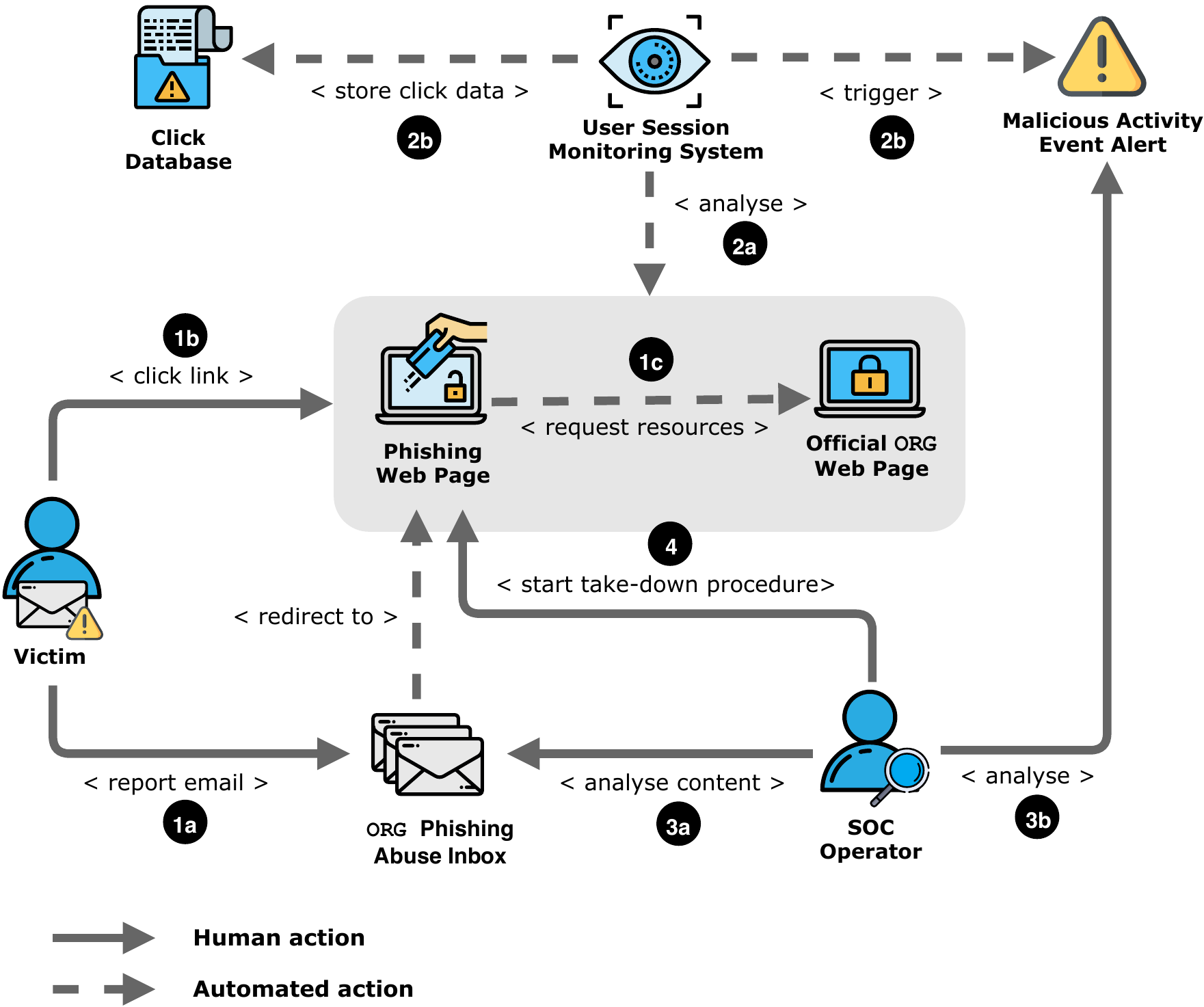}
\begin{minipage}{0.42\textwidth}
\footnotesize
\smallskip
SOC Operators collect evidence on the maliciousness of the web domain under investigation such that an external party can perform the notice and take-down requests for the malicious domains.
\end{minipage}
\caption{Overview of phishing-related activities at \ORG\ }
\label{fig:process}
\end{figure}

Our analysis relies on a unique dataset from a large phishing email database provided by \ORG, a large financial organization in Europe with more than 8 million customers and a multi-billion Euro turnover. 
\ORG\ customers that suspect they have received a phishing email in their personal email accounts are instructed by the organization to forward these emails to an internal \ORG\ functional mailbox. In parallel, \ORG's phishing response team runs a service to detect phishing domains (not necessarily linked
with the received phishing emails) by means of internal heuristics and limited to external domains requesting resources internal to \ORG\ (e.g. images, forms, logos, CSS files/javascript, etc.). {This data is generated by a third party service hired by \ORG\ that monitors \emph{all} requests generated towards \texttt{Org}'s resources}. Through this mechanism \ORG\ can detect the number of visits to the detected domains by accounting for the unique sessions opened between the (rogue) external and the (legitimate) internal services. Access to this data allows us to perform a rich analysis of the arrival of phishing emails, their characteristics, and to evaluate how often users have accessed malicious domains as a proxy measure of `phishing success'. Figure~\ref{fig:process} depicts \ORG's internal process to handle suspect phishing emails.

 Overall, we extracted 115,698 reported emails and 11,936 alerts for malicious links between February 1st, 2018 and 15 December 2018, with the exception of the period August-September 2018 due to infrastructural limitations at \ORG. For this same reason, our sample only includes data for `clicks' collected from end of July onwards. 
 
\paragraph{Data limitations {and ethical aspects}.} From the data structure, the link between a clicked URL and the specific email from which that click originated is not explicit and can only be reconstructed by exact match of the destination URL. 
This has the effect of limiting the scope of this study to the comparison of the effectiveness of cognitive influence techniques between phishing emails that are likely to have generated the click (as we cannot fully reproduce the process generating the detection of URLs that \emph{could} have been clicked, but have not). 
This also limits the number of matches between URLs reported in event alerts and URLs linked in emails. 
 {Further, the results of this work are limited to the emails that have been \emph{reported} (and therefore identified at least once) by \ORG's customers. Despite the large number of active reporting customers, particularly well-crafted emails may not be represented in our dataset. Further, we can only observe data captured by the \emph{User Session Monitoring System}, i.e. related to emails pointing to domains that `call back' to \ORG's systems. This may represent a limitation if emails that do not `call back' also exploit different `cognitive vulnerabilities', or with different distributions. However, an analysis on the available data does not show apparent biases between emails for which a `click' has been recorded, and those for which we do not know of any (ref. Figure~\ref{fig:barplotvulns}). These limitations are akin to those outlined by Pitsillidis \emph{et al.}~\cite{Pitsillidis2012}. Aware of these, we compensate by means of the analysis methodology that explicitly accounts for the potential biases in the data. Finally, the collected data did not contain sensitive subject information and all data handling has been performed within allowance from \ORG\ and within the scope of work previously approved by the department IRB.}

\subsection{Data sanitization and processing}
\label{sec:datasanit}

{As our email dataset contains messages forwarded by users, we first sanitize the data by removing} mobile text messages {($n=18,817$)} that likely result from erroneous forwards  to the functional mailbox from a related banking service; as they are irrelevant in our setting, we discarded them. 
{Further, users may have reported emails that target financial organizations different from \ORG.} To capture this, we identify targeted organizations in our dataset by a string search operation within email bodies for the names of the most prominent financial organizations in the country where \ORG\ is located, {and remove all records that do not belong to \ORG~($n=15,623$)}.
To identify phishing email subjects, dates, and recipient/sender information, we recursively searched through each raw email message to find header matches of the first original email arrived in the user's inbox,\footnote{{This is necessary as emails can be forwarded multiple times (e.g. if originally forwarded by the customer to an \ORG~employee) before ending up in the phishing inbox.}} and extract information on \texttt{From}, \texttt{To}, \texttt{Date}, and \texttt{Subject} values.
Table~\ref{tab:descstats} reports summary statistics of the final dataset.\footnote{
 {We notice that the upper 2.5\% of the distribution of email length is disproportionally long w.r.t. the remainder of the distribution, suggesting a few outliers in the data. Manual inspection reveals malformed email corpora (e.g. with HTML tags embedded in the body); as no obvious `upper limit' for email length is apparent, we keep these in the dataset for the sake of transparency.}}
\begin{table*}[t]
\centering
\footnotesize
\caption{Descriptive statistics of the collected dataset} 
\begin{minipage}{0.98\textwidth}
\footnotesize
The column \texttt{type} indicates whether the variable is a factor (f) or numeric (n). The column \texttt{n} reports number of levels for factors, and number of records with at least one observation for numerical variables. We do not report summary statistics for factors. The standard deviation for variable \texttt{Date} is reported in days. All dates are in 2018 and in format $\%m-\%d$. \smallskip
\end{minipage}
\label{tab:descstats}
\begin{tabular}{l|llllllllllllll}
  \toprule
 \multicolumn{1}{c}{} && & \multicolumn{6}{c}{Feb-Jul 2018} & \multicolumn{6}{c}{Oct-Dec 2018}\\
 \cmidrule(lr){4-9} \cmidrule(rl){10-15}
 \multicolumn{1}{c}{} &Variable& type & n & Min & 0.025q & Median & 0.975q & Max & 
 										n & Min & 0.025q & Median & 0.975q & Max \\ 
  \midrule

 \multicolumn{1}{c}{}& Language & f & 3 &  &  &  &  &  & 2 &  &  &  &  &  \\ 
 \multicolumn{1}{c}{}& To & f & 38760 &  &  &  &  &  & 2239 &  &  &  &  &  \\ 
 \multicolumn{1}{c}{}& From & f & 1641 &  &  &  &  &  & 330 &  &  &  &  &  \\ 
 \multicolumn{1}{c}{}& Date & n & 69800 & 02-02 & 03-07 & 05-30 & 07-28 & 07-31 & 11458 & 10-01 & 10-01 & 11-29 & 12-11 & 12-11 \\ 
 \multicolumn{1}{c}{}& Length & n & 69800 & 160 & 446 & 1068 & 3973 & 67246 & 11458 & 173 & 329 & 1320 & 5480 & 15685 \\ 
 \multirow{6}{*}{\rotatebox{90}{Vuln. triggers}}& Reciprocity & n & 69800 & 0 & 0 & 3 & 67 & 149 & 11458 & 0 & 0 & 2 & 37 & 153 \\ 
 & Consistency & n & 69800 & 0 & 0 & 13 & 84 & 132 & 11458 & 0 & 0 & 19 & 88 & 176 \\ 
  & Social Proof & n & 69800 & 0 & 0 & 2 & 17 & 52 & 11458 & 0 & 0 & 0 & 17 & 90 \\ 
  & Authority & n & 69800 & 0 & 0 & 5 & 55 & 121 & 11458 & 0 & 0 & 5 & 27 & 83 \\ 
  & Liking & n & 69800 & 0 & 0 & 0 & 7 & 504 & 11458 & 0 & 0 & 0 & 8 & 198 \\ 
  & Scarcity & n & 69800 & 0 & 1 & 40 & 107 & 157 & 11458 & 0 & 0 & 11 & 91 & 189 \\ 
 \multicolumn{1}{c}{} & Spoof dist. & n & 61911 & 0 & 0 & 7 & 14 & 23 & 10604 & 0 & 0 & 6 & 14 & 24 \\ 
     \multicolumn{1}{c}{}& Clicks & n & 4 & 9 & 9 & 28.5 & 78 & 78 & 35 & 1 & 1 & 37 & 220 & 220 \\ 

      \multirow{4}{*}{\rotatebox{90}{Emails}}& Reported & f & 69800 &&&&&&11458 \\
  & \hspace{0.05in} of which susp. & f & \hspace{0.05in}61079 &&&&&&\hspace{0.05in}9419 \\
  & Unique & f & 1293 &&&&&&424 \\
  & \hspace{0.05in} of which susp. & f & \hspace{0.05in}952 &&&&&&\hspace{0.05in}329 \\
   \bottomrule
\end{tabular}
\end{table*}

\subsubsection{Identification of suspicious and landing URLs}

{\textbf{Suspicious URLs.}} We check emails for the presence of suspicious URLs that point to any domain that does not belong to \ORG, as these would not normally appear in a legitimate email originated by the organization. We exclude from the heuristic general-purpose domains with no direct phishing correlation (e.g. \url{youtube.com}). Based on this classification we flag emails that contain at least one suspicious URL as \texttt{Suspicious}, whereas the remaining ones are considered uninteresting within our scope (as we can neither count nor estimate clicks for URLs that do not exist). 

\noindent{\textbf{Landing URLs.} These are landing URLs that load resources internal to \ORG, as detected and reported by the \emph{User Session Monitoring System} (ref. Fig.\ref{fig:process}). Whereas they are related to a click on a suspicious URL, this relation is not immediate in the data and needs to be reconstructed.}

\subsubsection{Landing URL extraction}
\label{sec:suspiciousLinks}

{To reconstruct the association between Landing URLs 
and Suspicious URLs we adopt the following method:
\vspace{-0.05in}
\begin{enumerate}
\itemsep0em 
 \item 
First, we traverse the suspicious URL embedded in the phishing email (\texttt{suspiciousURL}) multiple times by visiting all URLs arriving to \ORG's inbox. These typically generate a number of redirections (generally HTTP 3xx) that lead to a landing webpage, where the actual phishing resource is located. We record the association $\langle$ \texttt{suspiciousURL, landingUrl} $\rangle $ for \emph{all} visited URLs, and for \emph{all} emails; if the redirection mechanism is not deterministic, we obtain a $1\ to\ n$ association between \texttt{suspiciousURL} and a set of \texttt{landingURL}s. As we cannot know how many `redirection chains' exist from a single \texttt{suspiciousURL}, we traverse the URL opportunistically every time it appears in \texttt{Org}'s inbox. To minimize confoundings in the redirection, each visit session is independent from the previous. Figure~\ref{fig:countvsredirections} 
\begin{figure}
\centering
	\includegraphics[width=0.48\columnwidth]{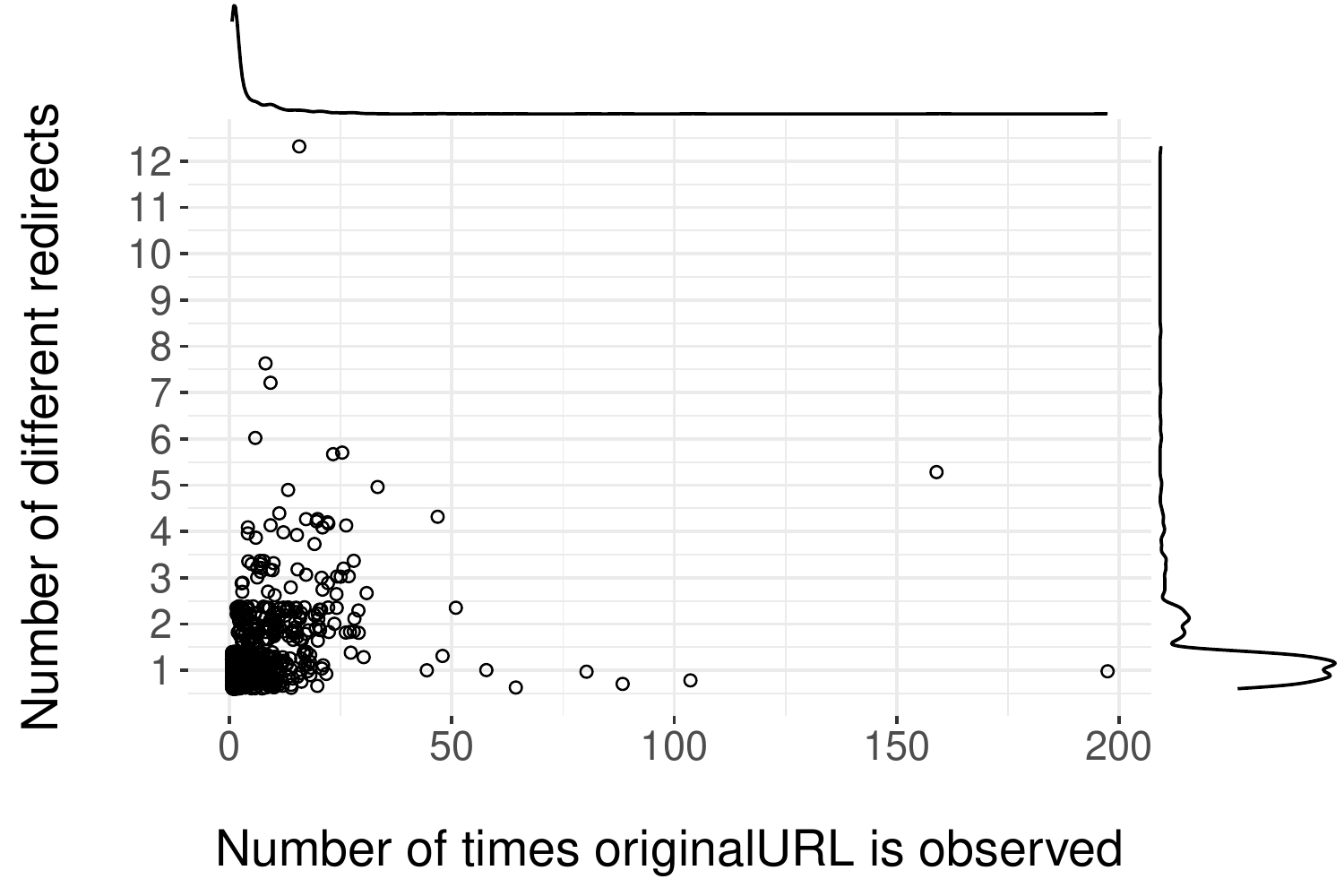}
	\includegraphics[width=0.48\columnwidth]{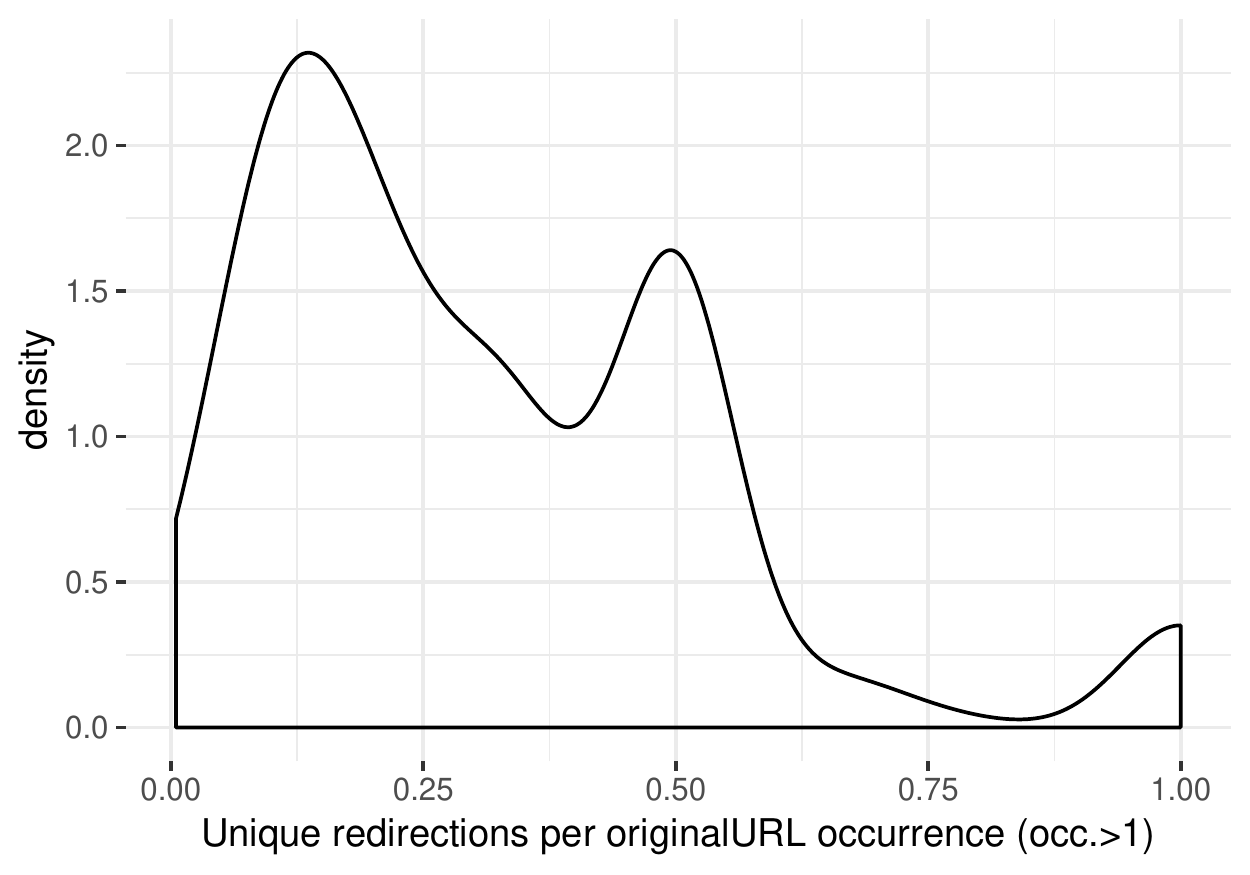}
\begin{minipage}{0.95\columnwidth}
\footnotesize
The redirection count for all observed \texttt{suspiciousURL}s (left) shows that new \texttt{landingURL}s stop appearing after only few \texttt{suspiciousURL} visits. The density plot on the right shows ratio of unique \texttt{landingURL} per \texttt{suspiciousURL} for \texttt{suspiciousURL}s visited more than once, and confirms that new redirects stop appearing regardless of number of visits.
\end{minipage}
\caption{Redirection count (left) and density ratio (right) from observed \texttt{suspiciousURL}s.}
\label{fig:countvsredirections}
\end{figure}
shows that the number of different redirections stops growing quickly regardless of how many time we traverse a given URL, suggesting that the dataset of collected \texttt{landingURL}s does not suffer from systematic censoring problems.

\item When \texttt{landingURL} is visited, a third party contractor of \texttt{Org} records a `click' for \texttt{landingURL} (see  Fig~\ref{fig:process} and discussion in \emph{data limitations}), and reports it to \ORG.

\item We link clicked \texttt{landingURL}s with the original email body by matching them with the \texttt{landingURL}s we found by traversing the \texttt{suspiciousURL}s in the mail corpus; if there are multiple clicked \texttt{landingURL}s for a single \texttt{suspiciousURL}, we keep record of all matches.

\item To aggregate clicks to a single \texttt{suspiciousURL}, we considered: average, sum, and max no. of clicks across all \texttt{landingURL}s for a given \texttt{suspiciousURL}. We ran our experiments using all aggregation strategies, and obtained qualitatively identical results. In this paper we report average clicks as it is the most conservative choice to make (e.g. summing \texttt{landingURL} clicks is more susceptible to over-reporting multiple clicks by the same user).
\end{enumerate}
}

{Figure~\ref{fig:urlmatches}
\begin{figure*}
\centering
\includegraphics[width=0.9\textwidth]{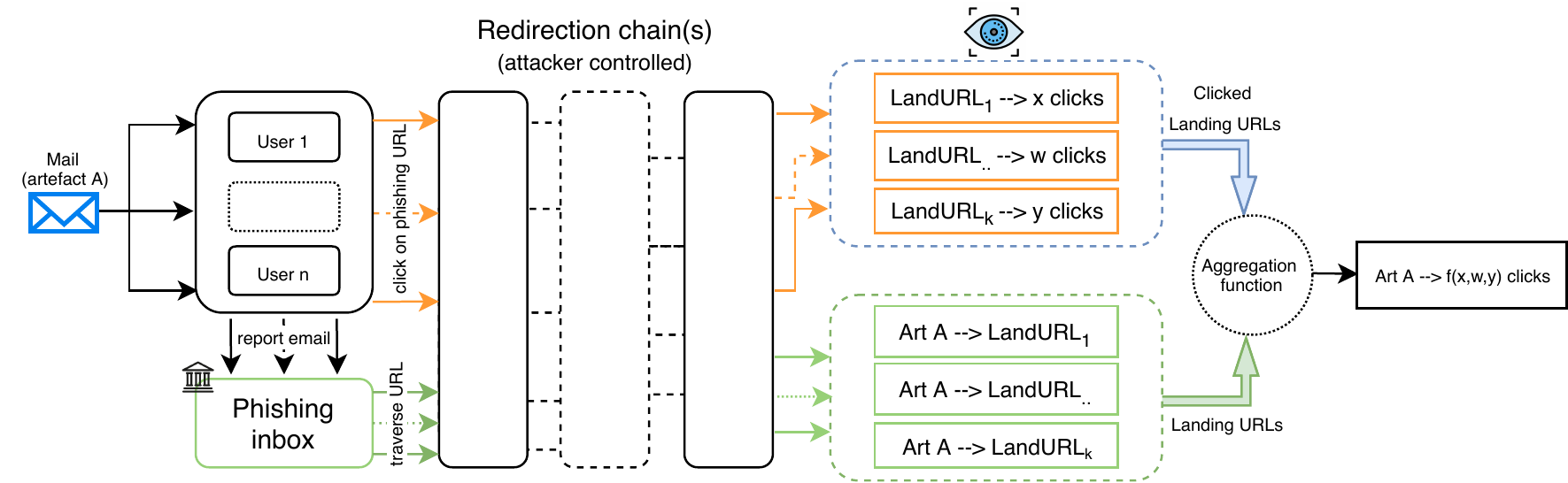}
\begin{minipage}{0.9\textwidth}
\footnotesize
 Data generation process for an example email leading to $k$ distinct clicked \texttt{landingURL}s. Users may click on the \texttt{suspiciousURL} link in the email and, usually through a series of redirections, reach the phishing domain hosted at one of the \texttt{landingURL}s. Through the dynamics described in Figure~\ref{fig:process}, an association between each distinct \texttt{landingURL} and recorded number of clicks is reported.
 The same email can also be reported to \ORG's phishing inbox. When it arrives, we opportunistically traverse the redirection chain and record the association between the original email and the final \texttt{landingURL}(s). To reconstruct the association between \texttt{suspiciousURL}s and clicked \texttt{landingURL}s, we aggregate the two datasets.
\end{minipage}
\caption{Data generation process for matched URLs and click data aggregation}
\label{fig:urlmatches}
\end{figure*}
provides a bird's eye view of the data generation process for the landing URL extraction.}

\subsubsection{Duplicate detection}
\label{sec:dupldetect}
One complexity of our unstructured dataset is the possible occurrence of multiple duplicates of the same suspect phishing email. In this paper we consider `similar' emails received by users over long periods of time as belonging to the same `campaign'.\footnote{This is only based on the email text, and we use it as a term to group together emails that are likely to have a common denominator (e.g. a phishing tool, a specific market/phishing pool, or actual attacker).} Although the overall textual content of these duplicate emails is similar, they can still contain slight differences, for instance because of the presence of a recipient's name in the salutation of an email or other minor syntactic features. In order to detect, and subsequently remove, as many of these duplicate emails as possible, we used a fuzzy string matching approach to determine the pairwise similarity for each of the emails in our dataset. 
We employ a \textit{bag-of-words} model to calculate, for each document, the frequency of each unique word in the document. We build the word-by-document matrix of our email corpora for the term frequency values for all emails in our dataset. As an additional pre-processing step all input was cleaned by removing special characters, urls, email addresses and line breaks from the text. We use $L^2$ normalization to the term frequencies to limit the impact of differences in email lengths \cite{Singhal2011}. 


To evaluate email similarity we employ a measure of cosine similarity. This similarity measure expresses the similarity between two vectors in terms of the cosine of the angle between the two vectors; the evaluation results in a score between $[0,1]$, where $0$ constitutes low similarity, and $1§$ constitutes high similarity. To define the cutoff threshold for similar emails we manually marked 300 randomly sampled emails from the dataset and assigned them to `similarity IDs' to track which emails were replicas of which others. We then performed a bootstrapped ($n=100,000$) sensitivity analysis of the threshold level to determine the optimal level for the cutoff. This procedure tunes the categorization to very satisfactory sensitivity and specificity levels higher than 90\%. Full details on procedure and results are reported in the Appendix.

{The duplicate detection procedure identifies $1,293$ and $424$ unique emails in the data collection of Feb-Jul 2018 and Oct-Dec 2018 respectively (ref. Table~\ref{tab:descstats}). Of these 952 and 329 respectively are classified as `suspicious'.\footnote{Note that otherwise identical emails may lead to different phishing domains.}}

\subsection{Cognitive evaluation}
\label{sec:cognevalmethod}

\begin{table}[t]
\centering
\caption{Topic model performance results}
\label{table:LLDA}
\footnotesize
\begin{minipage}{0.9\columnwidth}
\footnotesize
\smallskip
We perform LLDA using Gibbs sampling iterations for parameter estimation and inference initialised with hyper parameters $\alpha = 1.0$, $\beta = 0.001$, $k_{labels} = 6$ and $N_{iterations} = 1000$.
\end{minipage}

\smallskip
\begin{tabular*}{\columnwidth}{@{\extracolsep{\fill}}lll}
\toprule
\multicolumn{1}{l}{} & Macro (sd)    & Micro (sd)    \\ \midrule
Sensitivity          & 0.709 ($\pm$0.016) & 0.807 ($\pm$0.016) \\
Specificity          & 0.714 ($\pm$0.042) & 0.813 ($\pm$0.038) \\
Precision            & 0.718 ($\pm$0.025) & 0.755 ($\pm$0.024) \\
F1                   & 0.725 ($\pm$0.020) & 0.760 ($\pm$0.020) \\ \bottomrule
\end{tabular*}
\end{table}
{To identify the presence of cognitive vulnerabilities in email bodies and the \emph{intensity} of the employed cognitive attacks,} we construct a supervised topic model based on Labeled LDA~\cite{Lin2011} (LLDA). LLDA models each input document as a mixture of topics inferred from labeled input data and outputs probabilistic estimates of label-document distributions, i.e $P(label_t|document_m)$, and word counts of label-specific triggers for each input document. In our application the labels correspond to Cialdini's principles of influence, detailed in Table~\ref{table:phishing example}, whereas documents correspond to the emails.

{For model training, we randomly sampled 99 emails (38 with clicks and 61 suspicious) out of the set of unique and suspicious emails in the dataset ($n=1,281$),\footnote{To have an indication of the effect of sample size on model performance, we first ran the training on 70 emails and added 29 (+40\%) at a second time, obtaining virtually identical results. 
To rule out sampling issues, we also performed a cross-validation procedure (reported) which suggested stable results.
Finally, manual checks on a random sample from the dataset of predicted labels found no obvious miscategorization.} and manually labelled them for presence of cognitive vulnerabilities. Due to language restrictions, we adopted a mixed approach whereby one author performed the labelling on the original data, and the second author blindly re-performed the labelling on an automatically-translated random sample (20 emails) of the labelled data.}
To assess model performance we performed a 5 times repeated 5-fold cross validation over the data.
Numerous approaches exist to evaluate the performance of multilabel classification problems like ours. Following \cite{Rubin2012}, we consider our problem as a label-pivoted binary classification problem, where the aim is to generate for each label strict yes/no predictions based on the document ranking for that label. For each label, we sort on the per document prediction values, and use the \texttt{PROPORTIONAL} method \cite{Rubin2012, Furnkranz2008} to define a rank-cutoff value that determines the top $N$ ranked items that will receive a positive prediction. For each label, we set $TOPN_i$ equal to the expected number of positive predictions based on training-data frequencies: For label $l_i$, $TOPN_i = ceil\left(\frac{N^d_{test}}{N^d_{train}} * N^{train}_i\right)$ where $N^d_{train}$ and $N^d_{test}$ refer to the total number of training and testing documents and $N^{train}_i$ is the number of training documents assigned label $l_i$. 

We have aggregated the performance results of our topic model using the \texttt{PROPORTIONAL} rank-cutoff method in Table~\ref{table:LLDA}. Unlike other rank-cutoff methods, this approach relies solely on labeling information from the training set, which makes it appropriate for use in real-world production settings as well. We report both macro scores ({averages computed over each result of the cross-validation procedure}), and micro scores ({computed over the aggregate of all cross-validation results}).
The obtained scores indicate a satisfactory fit over both projections. A manual analysis on randomly sampled emails  confirms that the procedure  appropriately assigns `topics' to emails. 
The final model is trained on the complete set of 99 labeled training documents that were previously used in cross-validation, and then applied to the unseen and unlabeled remainder of the full dataset. 
 Standard text cleaning procedures have been applied for removal of special characters and stop-words, sentence tokenization, and word stemming.

 {In this paper we refer to the `topics' assigned by LLDA to an email as the \textbf{cognitive vulnerabilities} exploited in that text, and to the words associated with that topic and present in the text as the \textbf{vulnerability triggers} for that cognitive vulnerability. With this we aim at distinguishing the \emph{presence} of a cognitive attack from its \emph{intensity} in the email text.}

 \paragraph{Example of training results}
 
 { We report below an example of a phishing email (translated to English) and its association with different cognitive vulnerabilities. We have indicated the relevant vulnerability triggers in \textit{italics} and refer to (1) \texttt{Liking}, (2) \texttt{Consistency}, (3) \texttt{Authority}, (4) \texttt{Social Proof}, (5) \texttt{Reciprocity} and (6) \texttt{Scarcity}:

\begin{myquote}{0.25in}
\small

\textit{(1) As a valued customer of \ORG\ }we always want to inform you of the latest updates and innovations in our system. We have recently switched to a new system that requires \textit{(4) all current customers} to replace their \textit{(2) current debit cards} by our newly-produced ones.

In connection with the new changes to the \textit{(3) European Safety Regulations}, \ORG\ wishes to alert all its customers to the availability of the new and improved debit cards that adhere to all \textit{(3) environmental and safety regulations}.

\textit{(1) \ORG\ strives to be environmentally friendly}. Therefore, our service team will recycle all current debit cards by mounting your \textit{(2) current AES Encryption Chip} on your renewed biological RFID payment card. For this reason, all current payment cards must be replaced. \textit{(5) By participating in our recycling program, the new debit card can be requested free of charge}. \textit{(6) After October 19th, 2018, a direct debit will be charged}.


\end{myquote}

From the example we can observe that the different cognitive vulnerabilities often appear alongside each other, and that a single vulnerability can even occur multiple times within an email body. Table~\ref{tab:exampleVulns} reports an excerpt of the classification results for the above message, and the learned keywords (translated in English) for each topic.\footnote{As the original text is not in English, to provide an accurate translation we report keyword matches for an example.}}
\begin{table}
\centering
\footnotesize
\caption{Example of extracted keywords for each topic}
\label{tab:exampleVulns}
\begin{tabular}{ll ll ll }
\toprule

\multicolumn{2}{c}{Reciprocity} & \multicolumn{2}{c}{Consistency} & \multicolumn{2}{c}{Social Proof} \\

\toprule
Word & p & Word & p &Word & p \\
\midrule

free & 0.024  & update & 0.026 &all & 0.035\\
participate & 0.016 & improve & 0.024 & customer & 0.011 \\
program & 0.011 & recycle & 0.018 & current & 0.005 \\
request & 0.010 & renew & 0.015 & require & 0.004\\

\toprule
\multicolumn{2}{c}{Authority} & \multicolumn{2}{c}{Liking} & \multicolumn{2}{c}{Scarcity} \\
\toprule
Word & p & Word & p &Word & p \\
\midrule

safety & 0.017 & valued & 0.022 & after & 0.031\\
regulate & 0.013 & friendly & 0.012 & charge & 0.027 \\
european & 0.010 & strive & 0.008 & direct & 0.020 \\
must & 0.007 & environment  & 0.005 & debit & 0.019\\

\bottomrule
\end{tabular}
\end{table}

\section{Exploratory analysis}
\label{sec:expl}
In this section we provide an exploratory analysis of the obtained email data set reported in Table~\ref{tab:descstats}.

\begin{figure}
\centering
\includegraphics[width=0.7\columnwidth]{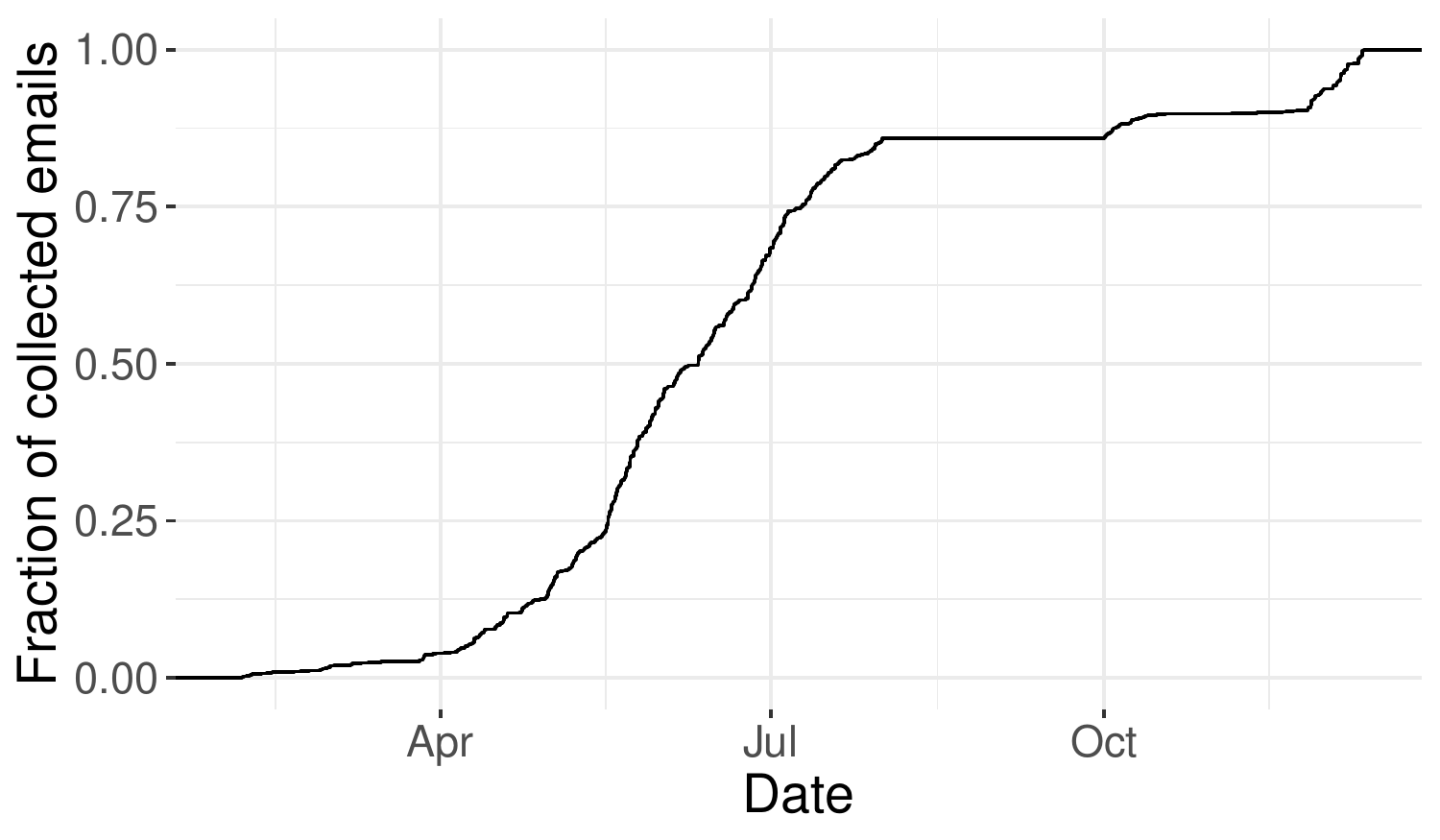}
\caption{Arrival of notified emails to \ORG's inbox}
\label{fig:arrival}
\end{figure}

We first give a look at the time of suspicious email arrivals in victims' inboxes. Figure~\ref{fig:arrival}
reports the CDF distribution of email arrivals to \ORG's phishing inbox.
We observe a steady arrival rate through April and the first cutoff date in July 2018, suggesting that email arrival is approximately constant and uniformly distributed in time. As per the time of day of their arrival (not depicted here for brevity) 
we observe that few suspicious emails arrive in the users' inboxes during the weekend, with most phishing activity happening during the working days. This may suggest a strategic aspect of these campaigns aimed at increasing the credibility of the email source. On this same line, we find that most emails arrive between 9am and 5pm (business hours), and most arriving between 9am and 11am. Interestingly, these findings are all in line with optimal email send days and times for newsletters as reported by analyses from multiple popular online email marketing services \cite{Mailchimp2014, CampaignMonitor2014, SendInBlue2017, Propeller}, and is an indication that attackers may follow similar strategies.

\subsection{Spoofing and victimization}
\begin{figure}[t]
\centering
\includegraphics[width=0.7\columnwidth]{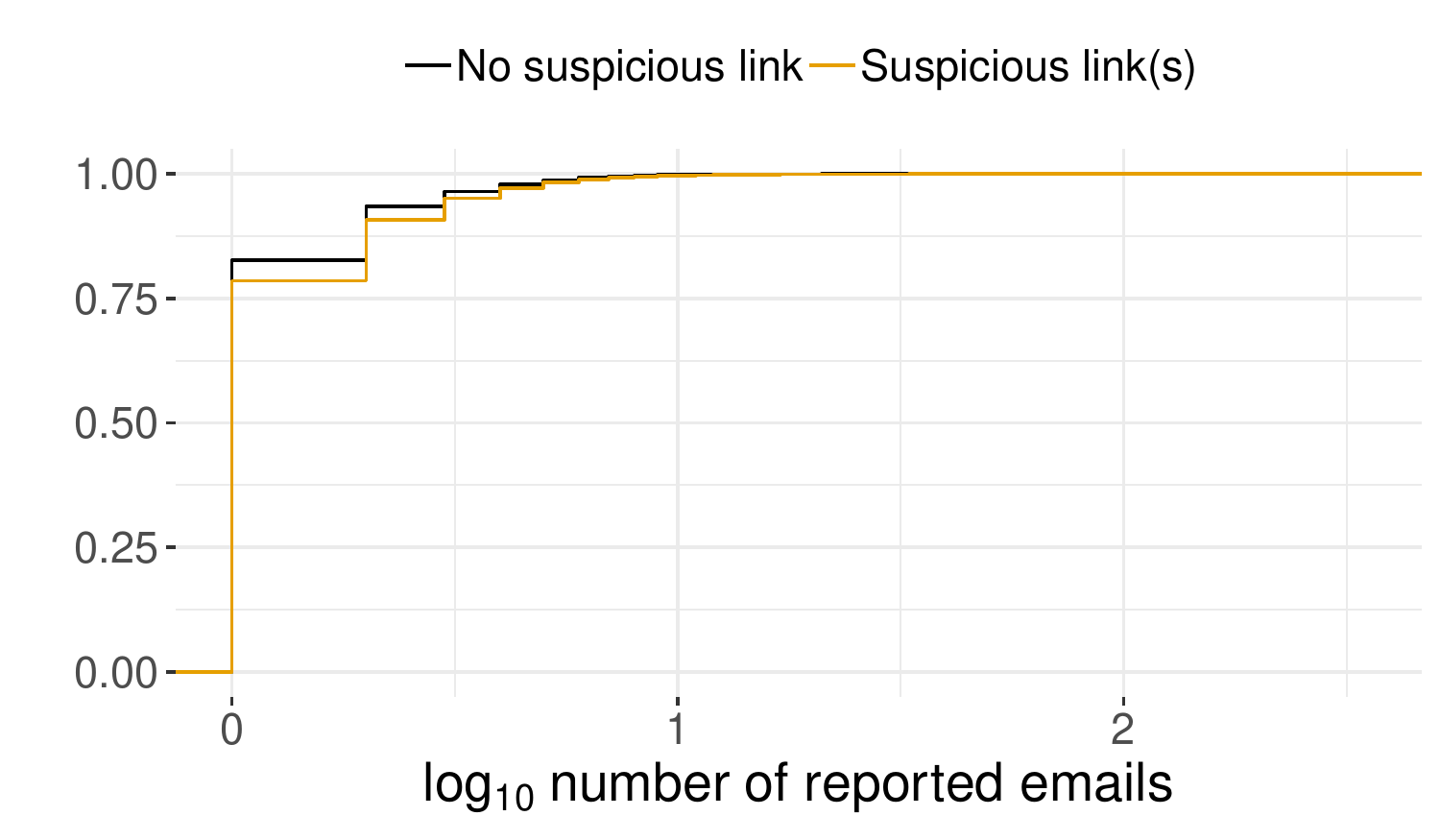}
\caption{CDF of emails reported by victim addresses}
\label{fig:cdftos}
\end{figure}

Figure~\ref{fig:cdftos} depicts the distribution of suspicious and
non suspicious reported emails.
The CDF is on a log scale to better represent the distribution's log tail. 
The vast majority of users report only one email, with almost all
reporting less than 10 emails. This suggests that the distribution of 
phishing emails is uniform across victims, as is generally the case with untargeted phishing attacks~\cite{le2014look,PhishLabs2018}. Only 122 addresses
out of about 40 thousand report more than 10 emails, and only nine 
report more than 100 emails.

\begin{figure}[t]
\centering
\includegraphics[width=0.75\columnwidth]{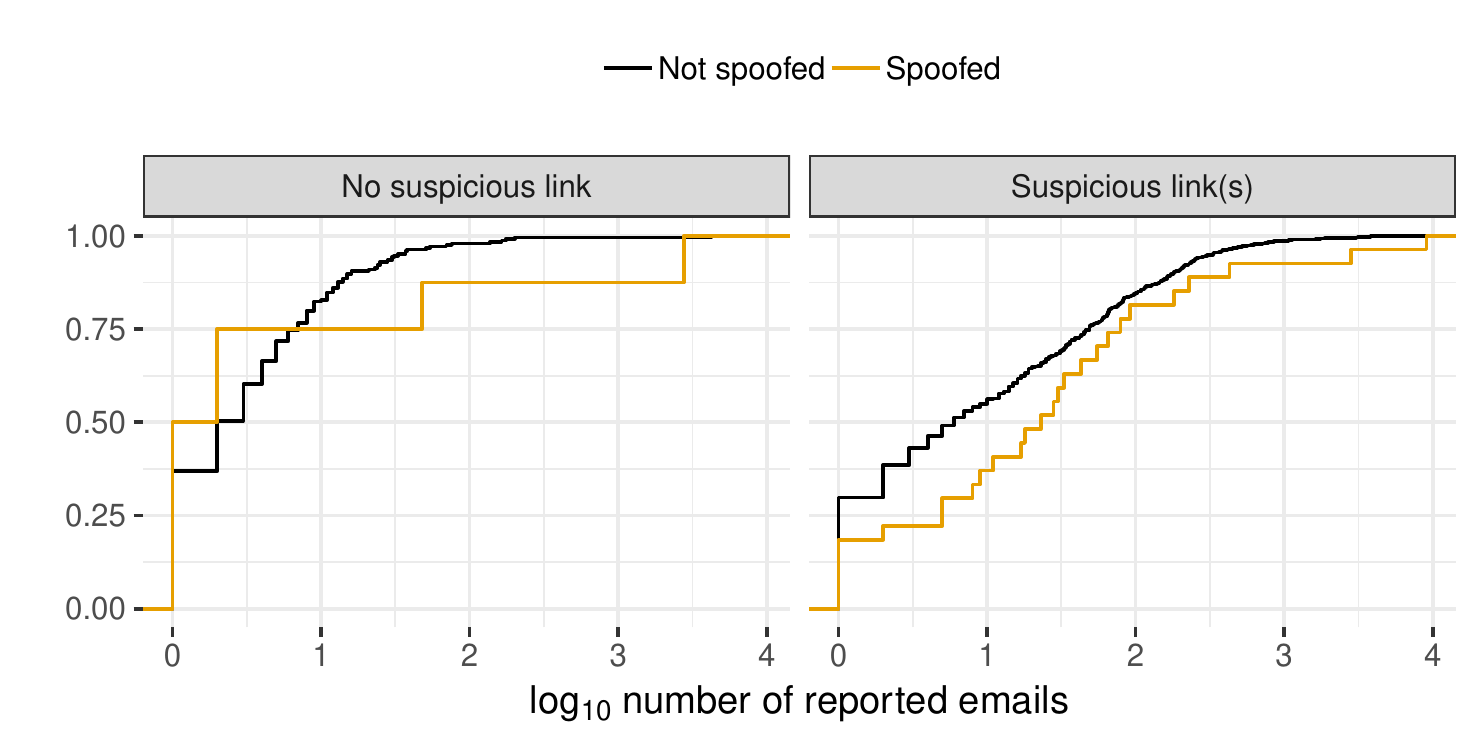}
\caption{CDF of spoofed and non-spoofed \texttt{From:} domains}
\label{fig:cdffroms}
\end{figure}

Figure~\ref{fig:cdffroms}
reports the distribution of spoofed and non-spoofed \texttt{From:} domains
for reported emails with and without a suspicious URL in the body. 
An email is classified as spoofed based on the Levenshtein distance 
of the (spoofed) \texttt{From:} domain the original attack was sent to, w.r.t. the actual name of the organization. This captures
exact string matches as well as small variations that may remain undetected by the user~\cite{Szurdi2014}. {We find attacks employing a range of domains resembling \ORG's: from less similar (e.g. \url{org-safety.com}, \url{org-customersupport.com}), to more closely spoofed domain variations (e.g. \url{theorg.com}, \url{0rg.com}).}
We observe
a clear differentiation, whereby emails with no suspicious URL are approximately as likely to have a spoofed \texttt{From:} address as a non-spoofed one. On the other hand, emails with suspicious URLs are more likely to be delivered from non-spoofed than from spoofed
addresses, {as can observed from the areas under the two curves}. This is compatible with a model of a relatively unsophisticated attacker. Here it is also relevant to consider that the pool of `spoofed' addresses is much smaller than the pool of `non-spoofed' addresses (as there are many fewer viable choices similar to \ORG\ than otherwise), suggesting that as spoofed domains get blacklisted, attackers may be forced to move to less well-spoofed \texttt{From:} addresses. 

\subsection{Phishing campaigns}

\begin{figure}[t]
\centering
\includegraphics[width=0.75\columnwidth, trim={0 4.5cm 0 5cm}, clip]{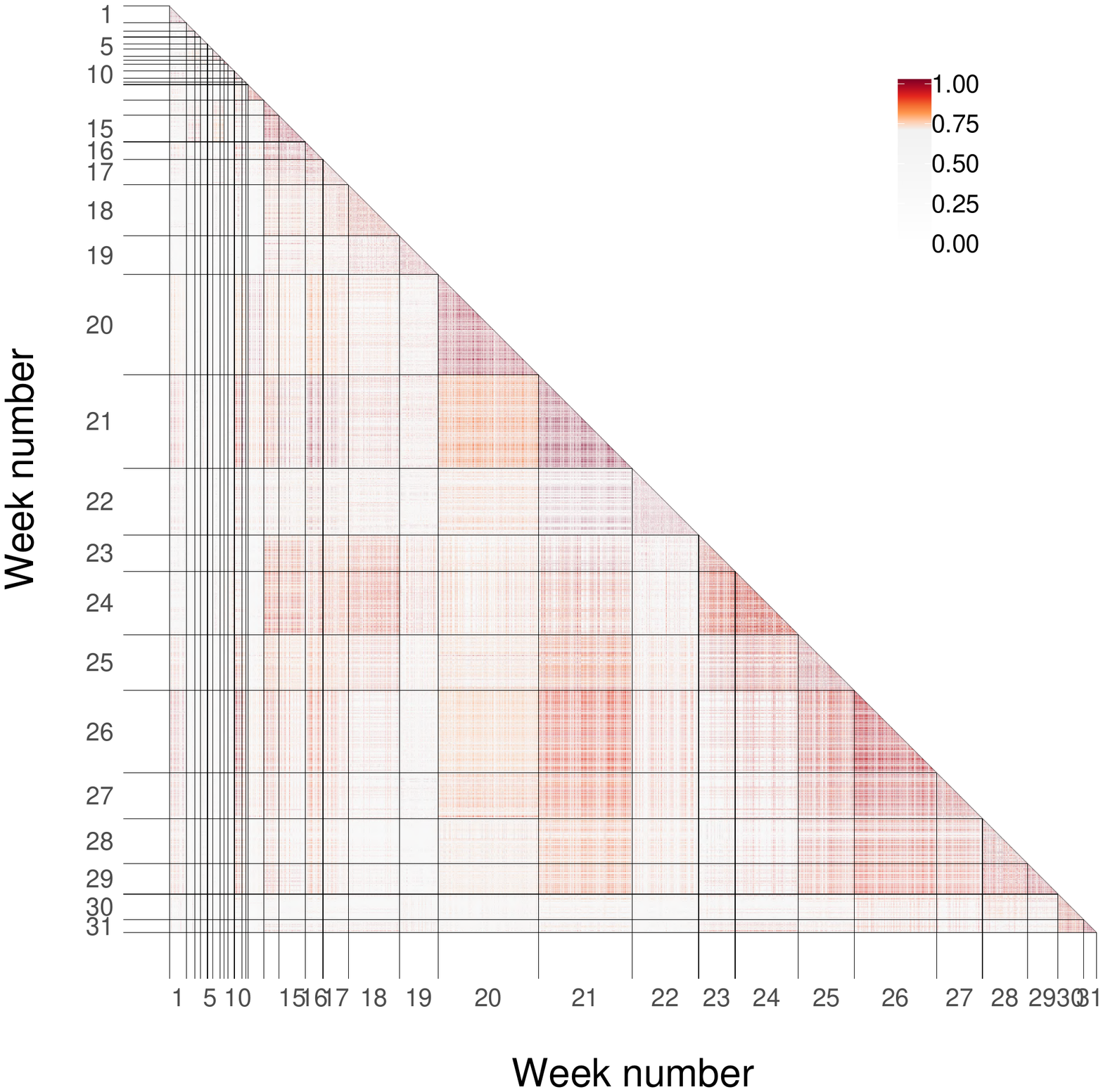}
\begin{minipage}{0.9\columnwidth}
\smallskip
\footnotesize
For visualization purposes we report random samples per week of 10\% of the emails received in that week. Red represent high similaries above the threshold. We do not observe specific cycles of similar emails, suggesting that any sufficiently long period of time (3-4 weeks) would cover a diverse set of phishing attacks.
\end{minipage}
\caption{Pair-wise cosine similarity between email samples}
\label{fig:simmatrix}
\end{figure}

\begin{figure}[t]
\centering
\includegraphics[width=0.7\columnwidth]{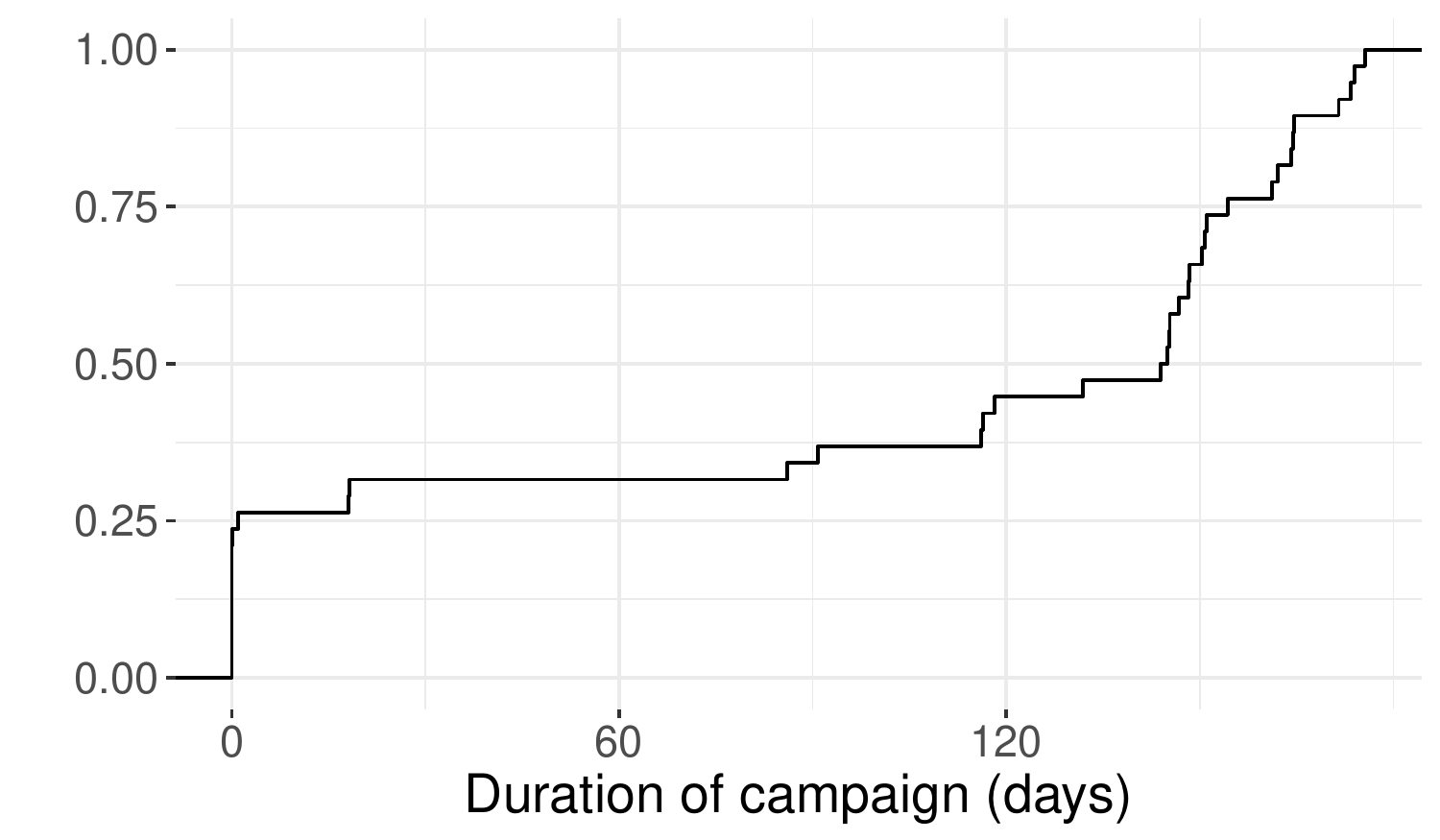}
\caption{Duration of a phishing campaign}
\label{fig:duration}
\end{figure}

Figure~\ref{fig:simmatrix}
reports a visualization of the similarity scores between emails received during the observation period. Dark red indicates high similarity.\footnote{For details on the identification of similar emails see the Appendix.} We do not observe specific and systematic cycles of campaigns emerging with repeating patterns across several weeks. This also suggests that any sufficiently long
observation period (in the order of 3-4 weeks) may suffice to collect a diverse set of attacks for analysis. {A first look suggests that some attacks seem to re-appear after a few weeks in slightly different forms, perhaps to increase chances of passing updated spam filters (see for example emails from week 21 reappearing slightly modified in week 26, or those from week 18 reappearing in week 24).} To evaluate this,
Figure~\ref{fig:duration}
reports the distribution of suspicious emails that likely belong to the same campaign. {Most campaigns are relatively long, with approximately 50\% of similar emails arriving more than 120 days apart, and 25\% of emails arriving more than 150 days apart with a relatively long left tail. From the distribution it appears that {\emph{single-day}} campaigns are relatively common, whereas long campaigns extend for more than 100 days. Mid-range campaigns lasting between 2 and 100 days are by comparison only few, suggesting that attacks may either be extremely quick and disappear the next day, or last for long periods.} Table~\ref{tab:campstats}
\begin{table*}[t]
\centering
\footnotesize
\caption{Descriptive statistics of duration and intensity of phishing campaigns} 
\label{tab:campstats}
\begin{minipage}{0.9\textwidth}
\smallskip
\footnotesize
\texttt{{SINGLE-DAY}} campaigns last up to one day; \texttt{SHORT} campaigns up to 100 days; \texttt{LONG} campaigns more than 100 days. Most phishing campaigns are either very short (one day) or long, with only a handful lasting more than one day but less than 100.
\smallskip
\end{minipage}
\begin{tabular}{lrrrrrrrrrrrrrrr}
  \toprule
&& \multicolumn{7}{c}{Phishing samples {(\#reported emails)}}&\multicolumn{7}{c}{Campaign duration (days)}\\
\cmidrule(lr){3-9} \cmidrule(lr){10-16}
 Type & n & Min & 1stQ & Mean & Med & 3rdQ & Max & sd & Min & 1stQ & Mean & Med &3rdQ & Max & sd \\ 
  \midrule
\texttt{SING.} & 10 & 1 & 1.0 & 1.3 & 1.0 & 1.0 & 3 & 0.7 & 0.0 & 0.0 & 0.1 & 0.0 & 0.0 & 1.0 & 0.3 \\ 
\texttt{SHORT} & 4 & 2 & 2.0 & 36.0 & 3.0 & 37.0 & 136 & 66.7 & 18.1 & 18.2 & 53.3 & 52.1 & 87.2 & 90.8 & 40.6 \\
\texttt{LONG} & 24 & 46 & 86.2 & 783.4 & 226.5 & 929.5 & 4827 & 1207.5 & 116.1 & 145.2 & 150.9 & 150.6 & 164.2 & 175.6 & 17.3 \\ 
   \hline
\end{tabular}
\end{table*}
reports summary statistics of suspected phishing campaigns. We identify 38 distinct campaigns lasting on average 150 days (approx 5 months) and up to 175 days in the observation period.

\begin{figure}[t]
\centering
\includegraphics[width=0.7\columnwidth]{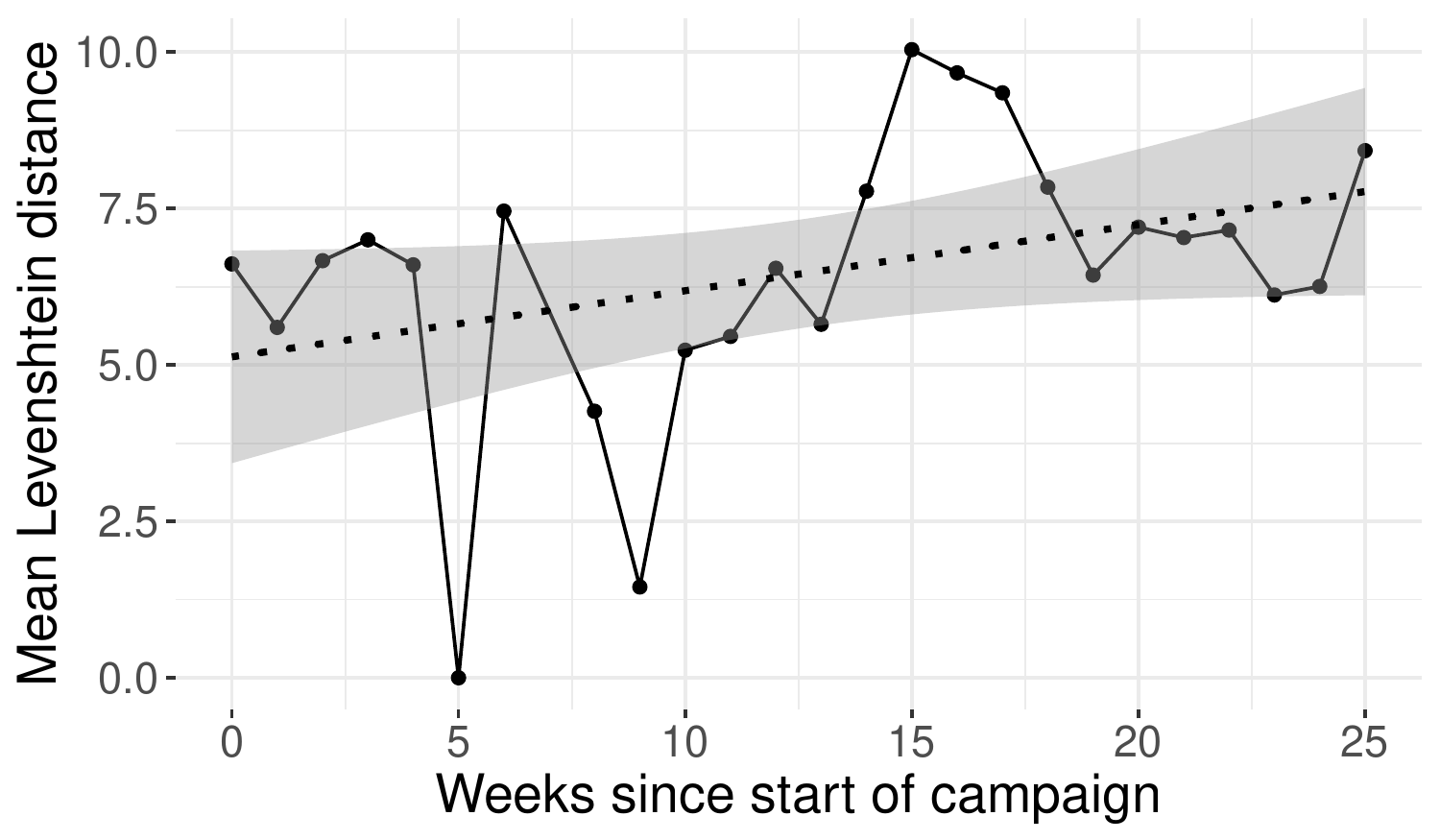}
\begin{minipage}{0.9\columnwidth}
\smallskip
\footnotesize
As phishing campaigns progress, the spoofed \texttt{From:} domains appear
to be more dissimilar w.r.t. the original domain.
\end{minipage}
\caption{Average weekly decrease in similarity between spoofed domains and name of target organization}
\label{fig:durationvsdistance}
\end{figure}

 To investigate how address spoofing evolves during campaigns, Figure~\ref{fig:durationvsdistance}
reports the weekly average similarity between the domain of the attacker \texttt{From:} address and the domain of the victim organization (measured as their Levenshtein distance) for \texttt{LONG} campaigns. Lower scores indicate more closely spoofed domains. We observe an average increase in dissimilarity between spoofed \texttt{From:} addresses and organization domain, which suggests an overall deterioration of a phishing campaign as it progresses or is replicated by phishers 
($cor=0.31, p=0.08$). {This is in line with the intuition that spoofed domains are limited in number, and attackers may therefore run out of options as domains get blacklisted as the campaign progresses.}

\subsection{Cognitive effects}
\label{sec:cogneval}


{Figure~\ref{fig:barplotvulns}
\begin{figure*}
\centering
\includegraphics[width=0.75\columnwidth]{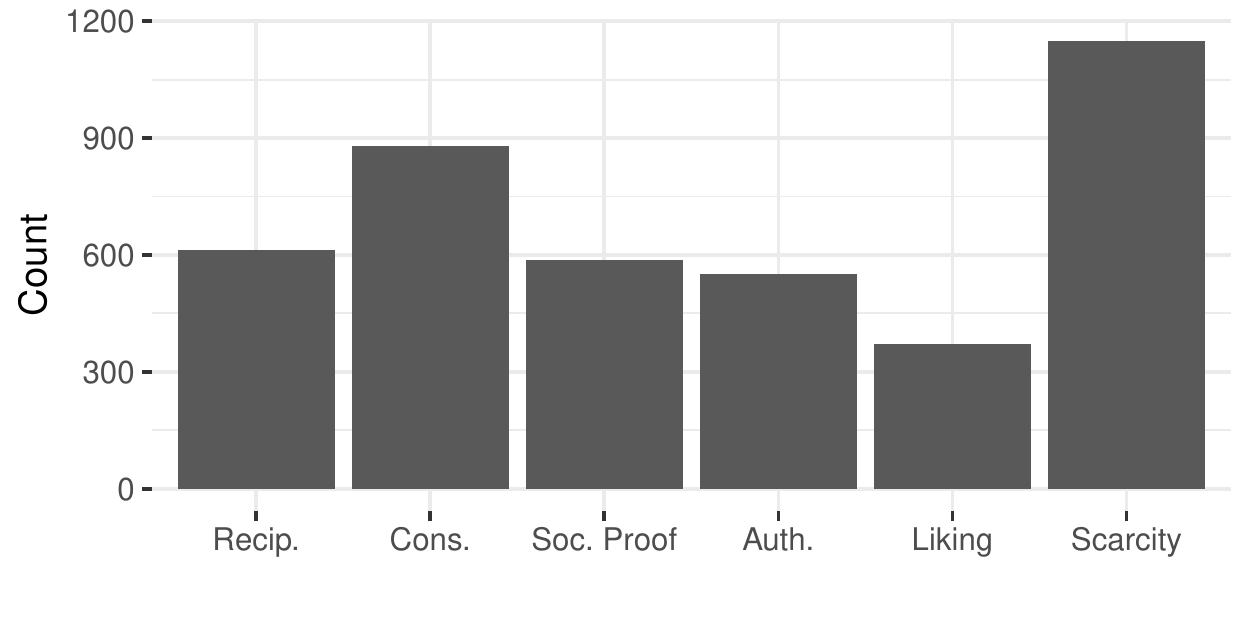}
\includegraphics[width=0.75\columnwidth]{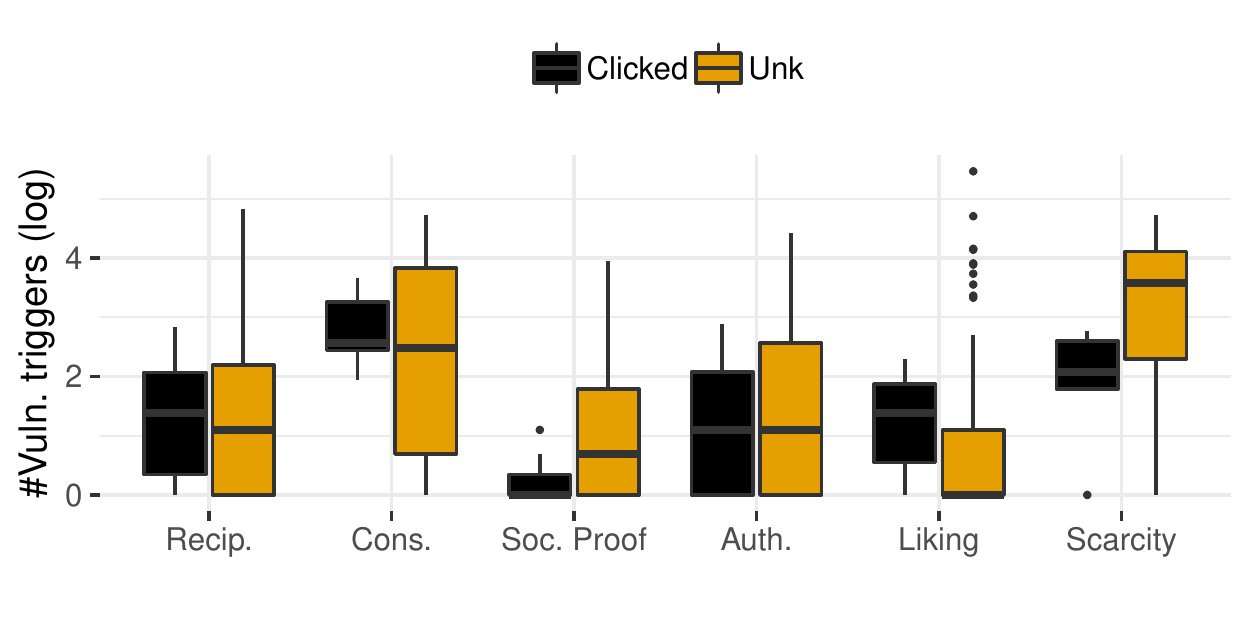}
\begin{minipage}{0.8\textwidth}
\footnotesize
Most phishing attempts trigger \texttt{Scarcity}, \texttt{Consistency}, and \texttt{Reciprocity} vulnerabilities.
\texttt{Social Proof}, \texttt{Authority}, and \texttt{Liking} are the least common. Relative frequency of cognitive
vulnerabilities is reflected in the distribution of vulnerability triggers identified in the emails. We do not identify specific biases in presence of vulnerability triggers between emails for which a `click' has been registerd, and emails for which it has not (i.e. that received an unknown number of clicks).
\end{minipage}
\caption{Distribution of triggered cognitive vulns. (left), and of vuln. triggers (right) for emails}
\label{fig:barplotvulns}
\end{figure*}
reports the distribution of triggered cognitive vulnerabilities in each unique email (left) and the corresponding vulnerability triggers identified in the corpus (right)}. We observe a clear relation between the two plots: the most common vulnerabilities and triggers in emails appear to be linked to the \texttt{Consistency} and \texttt{Scarcity} vulnerabilities, regardless of whether a `click' has been recorded for that link or not. \texttt{Liking} and \texttt{Social proof} triggers appear to be particularly rare on the average, with most emails targeting none.\footnote{{The descriptive statistics reported in Table~\ref{tab:descstats} also suggest stable distributions between the collection periods; for \texttt{Liking} we observe more extreme values (upper $97.5\%=7,\ max=504$ in the Feb-Jul data collection); this is caused by the outliers in the email corpora for which we measure disproportionate email lengths.}} This is consistent with the intuition that in one-shot interactions (as opposed to prolonged or repeated exchanges as in spear-phishing attacks~\cite{le2014look}) cognitive attacks linked to the target's social context and personal preferences (ref. Table~\ref{table:phishing example}) are rare. By contrast, exploiting \texttt{Consistency} may only require reference to previous actions that the group of potential victims will have likely performed, such as buying an insurance or receiving a debit card from the organization. \texttt{Authority} appears to be a relatively common trigger in our sample, albeit not for all emails. Common triggers here refer to European and national-level legislation and often come together with the threat of a punishment if certain actions are not completed. {Overall, we find that few cognitive triggers are present in the median email, suggesting that the median reported attack may not be highly effective, whereas few emails embed more `intense' cognitive attacks.}

\paragraph{Effect of cognitive vulnerablities on phishing success.}

\begin{figure}[t]
\centering
\includegraphics[width=0.8\columnwidth]{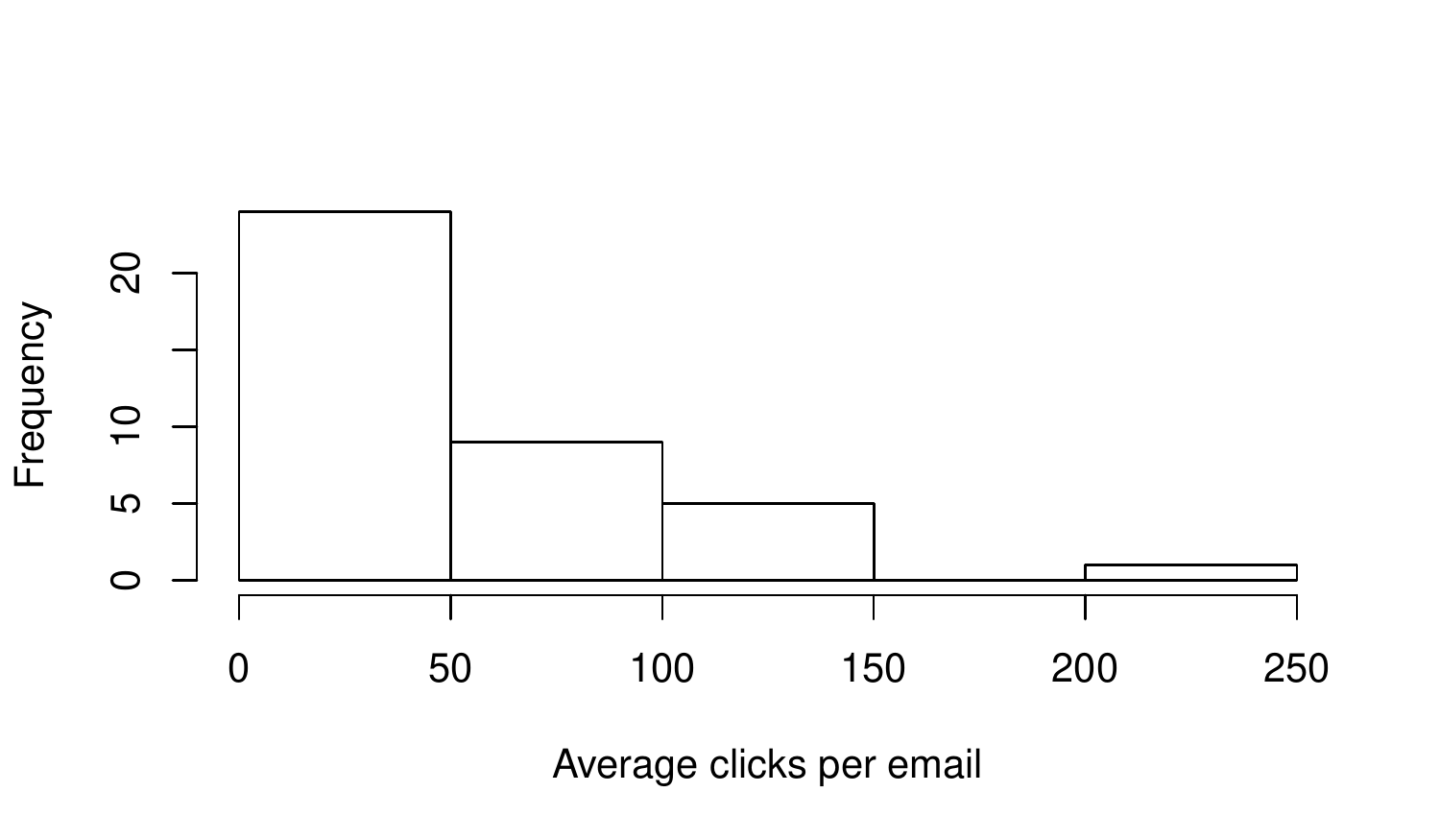}
\caption{Histogram distribution of clicks per email}
\label{fig:histclicks}
\end{figure}

\begin{figure}[t]
\centering
\includegraphics[width=0.7\columnwidth]{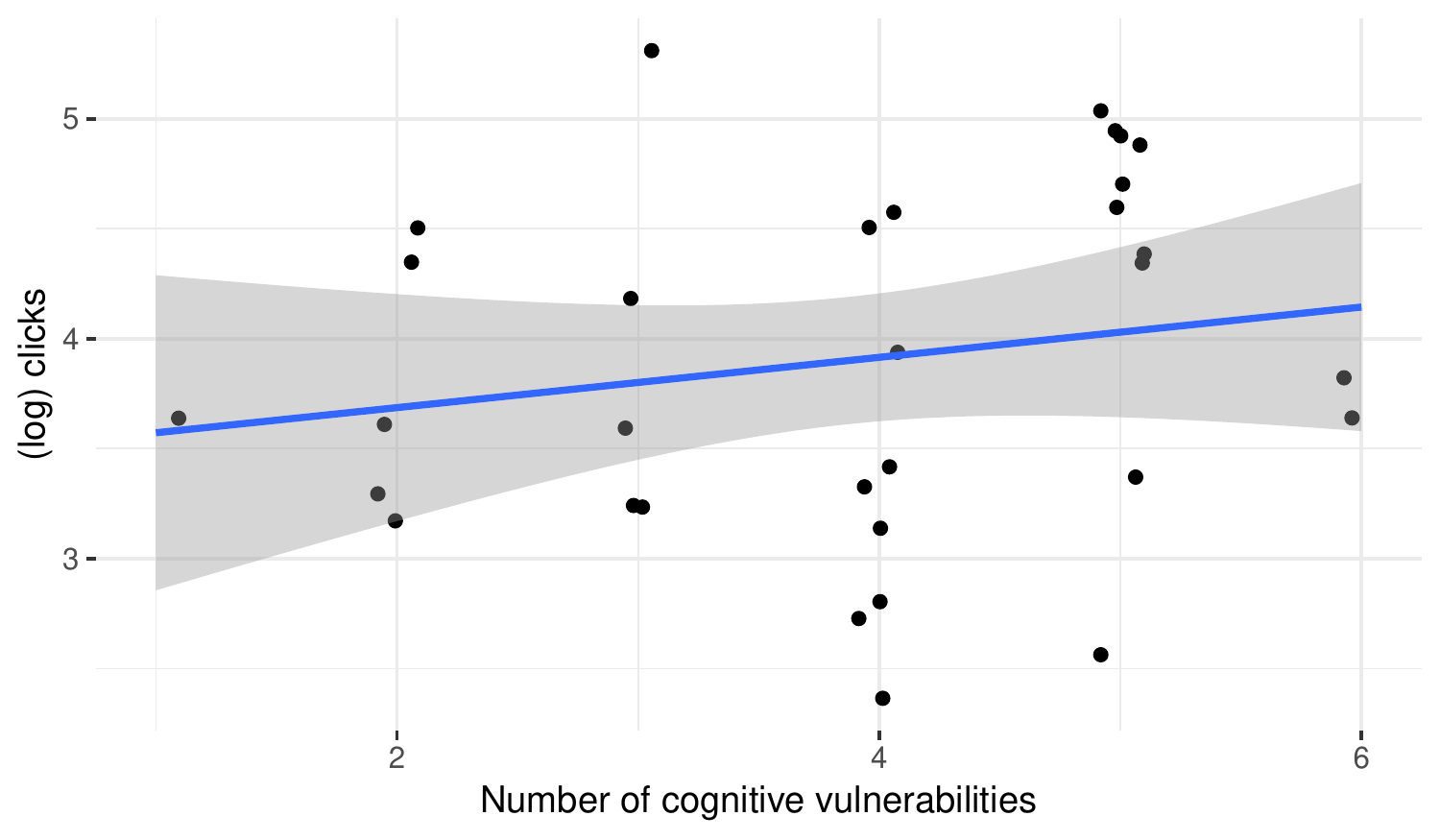}
\begin{minipage}{0.85\columnwidth}
\footnotesize
We observe a clear relation between the presence of exploited cognitive vulnerabilities and the clicks generated by the embedded URL(s). The shifted position of points in the pictures is to clear overlaps and is only presentational.
\end{minipage}
\caption{Relation between number of cognitive vulnerabilities in an email and average clicks ($log_{10}$)}
\label{fig:vulnsvsclicks}
\end{figure}

To evaluate the effect of the cognitive features of the email(s) embedding the `clicked' URL links, we first report in Figure~\ref{fig:histclicks} the distribution of average clicks generated by emails for which at least one click has been recorded {($n=40$)}.
Most emails generate fewer than 150 clicks, with two emails generating more than 200 clicks ($min=1,\ median=37,\ max=220,\ sd=51.9$).
Figure~\ref{fig:vulnsvsclicks}
displays the relation between triggered cognitive vulnerabilities and generated clicks, for which we observe a clear positive relation.\footnote{A possibility is that some emails may be distributed to substantially more users than other emails, generating greater aggregate click counts. As we have no access to the victim's inboxes, we cannot directly measure this. However, the data does not show specific biases in the likelihood of users reporting emails (Figure ~\ref{fig:cdftos}), suggesting that major skews are not realistic. This is consistent with previous findings in the literature~\cite{Yip2013,PhishLabs2018}. Further, due to the very low click-through rates of spam and phishing campaigns~\cite{Kanich-2008-CCS}, this difference should be of several orders of magnitude to have a visible effect (as opposed to be undetectable noise in the data generation process). {Regardless, in the Appendix we build a data generation model to evaluate the effect this bias would have in the data if present; our analysis finds no evidence.}}
Following common practice~\cite{lawless1987regression}, to avoid  dispersion we here only consider URLs clicked at least ten times, removing six emails. 
A simple Poisson regression of the form $log(clicks_i) = \alpha + \beta(cogvulns_i)$ reveals a strong positive correlation between the variables ($\beta=0.12, p<0.001$). This suggests that the more cognitive vulnerabilities are exploited in an email body, the more 
that email can be expected to generate compliant user behaviour, even when not considering the type of cognitive attack, or its intensity.

\paragraph{Effect of vulnerability triggers.}

\begin{figure*}[t]
\centering
\includegraphics[width=0.99\textwidth]{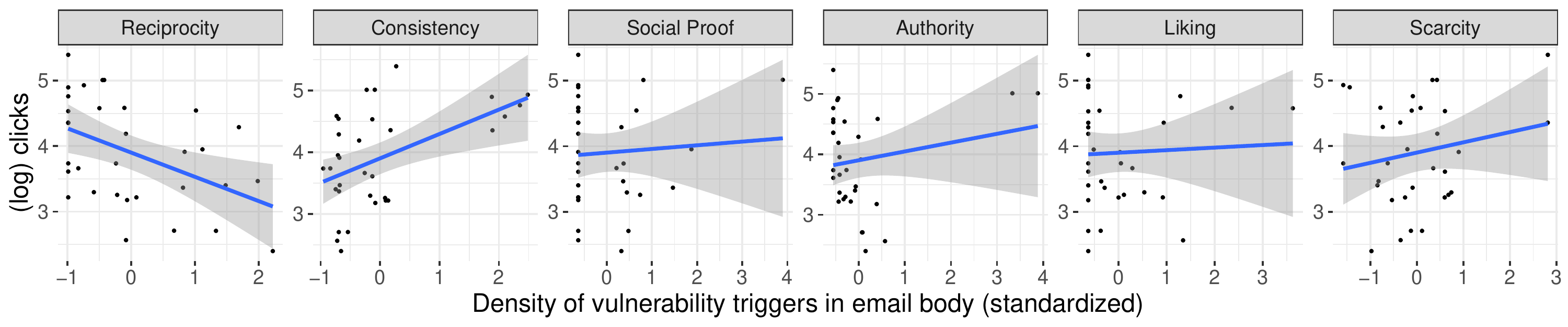}
\begin{minipage}{0.9\textwidth}
\footnotesize
The data shows the effect of different cognitive vulnerability triggers
on expected number of clicks. \texttt{Consistency} and \texttt{Scarcity} have a clear positive association with the expected number of clicks they generate. \texttt{Social proof}, \texttt{Authority} and \texttt{Liking} do not show any evident trend. Interestingly, we find that \texttt{Reciprocity} appears to be counterproductive.
\end{minipage}
\caption{Correlation between vulnerability triggers and observed clicks}
\label{fig:correlation}
\end{figure*}

We now consider the relation of the intensity of each cognitive attack (i.e. measured by the presence of vulnerability triggers) with the measured `success' of the phishing email. Figure~\ref{fig:correlation}
reports the results. The data reports a clear positive relation between \texttt{Consistency}, and \texttt{Scarcity} vulnerability triggers with the expected (log) number of clicks. \texttt{Reciprocity} shows a negative relationship. Additionally \texttt{Social proof}, \texttt{Liking} and \texttt{Authority} show no evident effect, whereby the majority of emails have relatively small counts of associated vulnerability triggers (see also Figure~\ref{fig:barplotvulns}). {On the other hand, looking at the right extreme of the scale, the few available data points are always related to highly-clicked emails; this may indicate that triggering these vulnerabilities (in this application domain) may be particularly difficult, for example as decisions related to personal finance may have a smaller attached  `social' component, or as adding additional `authoritative' effects in the banking domain may be challenging for an attacker.}

\begin{figure}[t]
\centering
 \includegraphics[width=0.7\columnwidth]{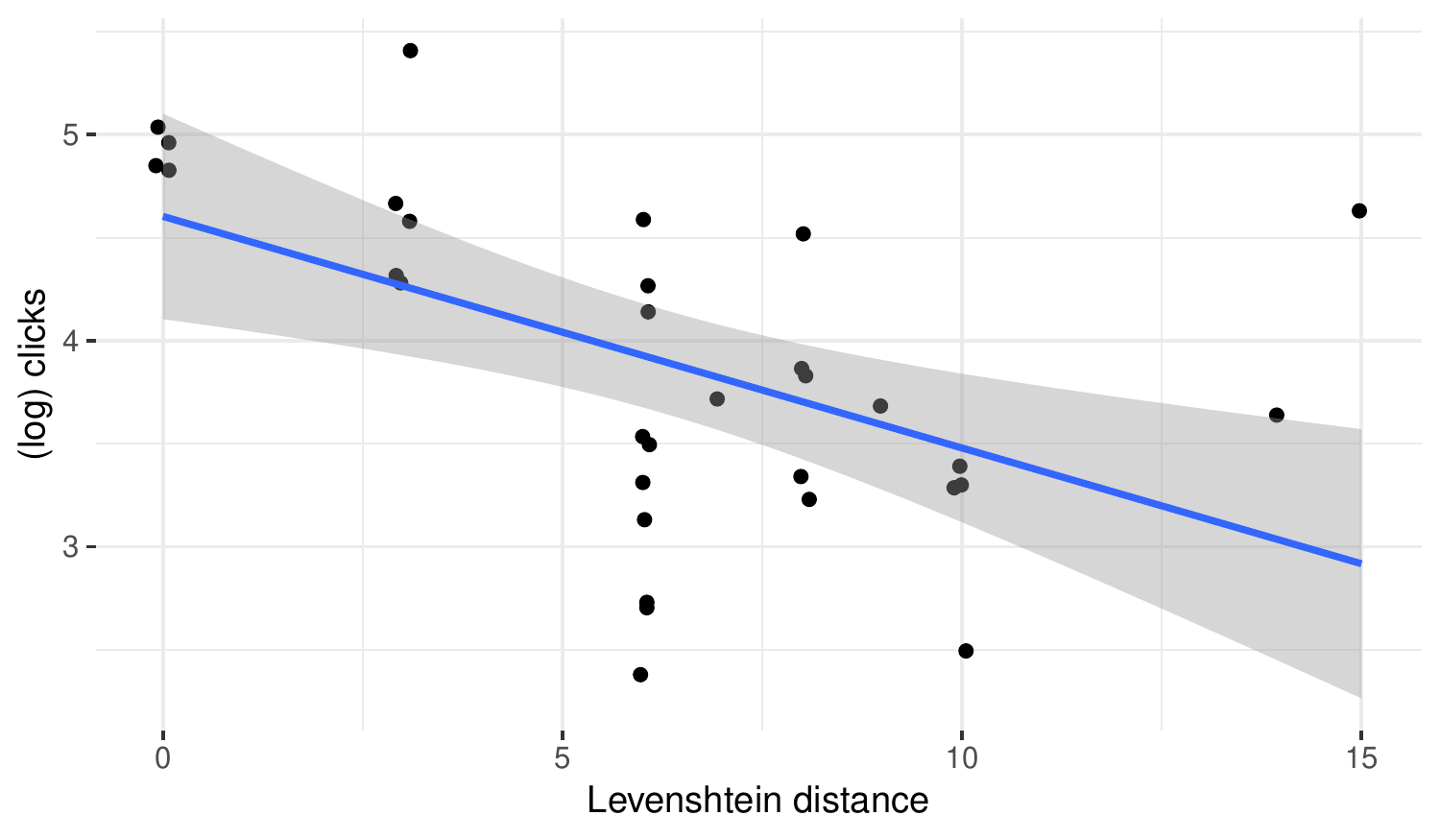}
 \begin{minipage}{0.85\columnwidth}
 \smallskip
\footnotesize
We identify a negative relation between the dissimilarity of the spoofed \texttt{From:} domain in an email against the original one, and the expected
number of clicks the email entices.
\end{minipage}
\caption{Relation between spoofing dissimilarity and average clicks ($log_{10}$)}
\label{fig:spoofvsclicks}
\end{figure}

\paragraph{Effect of spoofing distance.} Apart from the cognitive vulnerabilities exploited in the text, a second relevant factor could be the similarity between the \texttt{From:} address displayed to a user and \ORG's legitimate one.
Figure~\ref{fig:spoofvsclicks} reports the relation between Levenshtein distance of the spoofed \texttt{From:} domain and the expected number of clicks.
We find an inverse relation between the two variables, suggesting that the greater the dissimilarity between the spoofed and the original domain, the lower the average number of generated clicks ($\beta = -0.13, p<0.001$).
This suggests that both cognitive attacks and the degree of spoofing in an email may have an effect on the relative success of a phishing email and could be considered to build a triaging model for phishing emails.

\section{Modelling phishing success}
\label{sec:modelling}

We now evaluate the relative impact of each cognitive variable in the collected dataset. We estimate
coefficients for a Poisson process of the (aggregate) form:
\begin{equation}
log(clicks_i) = \alpha + \beta_1 cogvulns_i + \beta_2 spoofdist_i + \epsilon_i
\label{eq:reg}
\end{equation}
whereby, for each email $i$, $clicks$ represents the number of measured clicks, $cogvulns$ is the array of counts of the vulnerability triggers identified in the email body, and $spoofdist$ indicates the degree of (dis-)similarity between the spoofed \texttt{From:} address and the original \ORG\ domain. $\epsilon_i$ is the error term. To monitor and account for overfitting problems related to the few available datapoints, we combine a step analysis of each model (M1..M7) with regression bootstrapping to generate robust confidence intervals for the coefficient estimations. For model selection we report coefficients, 95\% confidence intervals, residual deviance, and Adjusted McFadden Pseudo-$R^2$, to reduce the statistical bias in the performance metrics for model selection.\footnote{Importantly, with this procedure we \emph{do not} aim at identifying a definitive model and coefficients to forecast phishing success: regardless of the amount of observations in the dataset, that would not be possible because the `click generation process' generating the observations necessarily varies from domain to domain (e.g. finance vs health), from organization to organization (e.g. national vs international), and from customer base to customer base (e.g. sensibility of application domain). Therefore, coefficient estimations out of this type of models cannot be `\emph{plug-and-play}' across organizations and domains and will require tuning before being applied in-house.}
Results are reported in Table~\ref{tab:reg}.

\begin{table*}[t]
\centering
\caption{Regression results for Eq. \ref{eq:reg}} 
\label{tab:reg}
\footnotesize
\begin{minipage}{0.87\textwidth}
\smallskip
All model coefficients estimations are relatively stable across the seven models. Coefficients for the Poisson models are presented with 95\% confidence intervals in parentheses. \texttt{Social proof} and \texttt{Spoof distance} of \texttt{From:} addresses appear to have the largest effects on predicted number of clicks. Higher spoof distances (i.e. higher dissimilarity between \texttt{From:} domain and original domain) result in a lower number of expected clicks.
  We only report coefficient significance (indicated by a $^\star$ for significance at the $0.1\%$ level) for the reader's reference;
however due to the relatively small sample size coefficient estimations should only be interpreted relative to each other as opposed to in absolute terms. Model power w.r.t. the baseline model is reported by the adjusted McFadden Pseudo-$R^2$; {a $\chi^2$ test is employed for model comparison ($^\star:\ p\leq0.001$; $^\dagger:\ 0.001<p\leq0.01$).} Standard model checks do not reveal issues or biases in the model fit. 
\smallskip
\end{minipage}
\begin{tabular}{lrrrrrrr}
\toprule
& \textbf{M1} & \textbf{M2} & \textbf{M3} & \textbf{M4} & \textbf{M5} & \textbf{M6} & \textbf{M7} \\
\cmidrule{2-8}
$\alpha$ & 4.38$^\star$ & 3.89$^\star$ & 3.79$^\star$ & 3.63$^\star$& 3.37$^\star$ & 3.37$^\star$  & 4.22$^\star$ \\ 
& (4.33, 4.42) & (3.81, 3.97) &  (3.71, 3.87) & (3.54, 3.73) & (3.17, 3.44) & (3.23, 3.51) & (4.02, 4.42) \\
\texttt{Reciprocity} & -0.02$^\star$  & -0.01$^\star$ & -0.02$^\star$ & -0.02$^\star$ & -0.02 & -0.02$^\star$ & -0.02$^\star$\\
& (-0.02, -0.02) & (-0.02, -0.01) & (-0.03, -0.02) & (-0.02, -0.01) & (-0.02, -0.01) & (-0.02, -0.01) & (-0.02, -0.01)\\
\texttt{Consistency} &  & 0.02$^\star$ & 0.02$^\star$ & 0.02$^\star$ & 0.03$^\star$ & 0.03$^\star$ & 0.01$^\star$ \\
& & (0.02, 0.02) & (0.02, 0.02) & (0.02, 0.02) & (0.02, 0.03) & (0.02, 0.03) & ( 0.01, 0.02)\\
\texttt{Social proof} &  &  & 0.14$^\star$ & 0.11$^\star$  & 0.04 & 0.04 & 0.10$^\star$\\
& & & (0.11, 0.16) & (0.08, 0.14) & (0.01, 0.08) & (0.01, 0.07) & (0.06, 0.13)\\
\texttt{Authority} &  &  &  & 0.01$^\star$ & 0.02$^\star$ & 0.02$^\star$  & 0.00 \\
& & & & (0.01, 0.02) & (0.02, 0.03) & (0.02, 0.02)& (0.00, 0.01)\\
\texttt{Scarcity} &  &  &  &  & 0.02$^\star$  & 0.02$^\star$ & 0.02$^\star$\\
& & & & & (0.02, 0.03) & (0.02, 0.03) & (0.01, 0.02)\\
\texttt{Liking} &  & &  &  &  & -0.02$^\star$  & 0.04$^\star$ \\
& & & & & & (-0.04, -0.01) & (0.02, 0.06)\\
\texttt{Spoof dist.}&  & &  &  &  &  & -0.10$^\star$ \\
& & & & & & & (-0.12, -0.08)\\
\midrule 
Adj. Pseudo-$R^2$ & 0.09 & 0.23 & 0.28 & 0.30 & 0.33 & 0.33 & 0.41 \\
{$Res.\ Dev.$} & 1390$^\star$ & 1136$^\star$ & 1054$^\star$ & 1012$^\star$ & 958$^\star$ & 951$^\dagger$ & 814$^\star$  \\
{N} & 38 & 38 & 38 & 38 & 38 & 38 & 38 \\ 
\bottomrule
\end{tabular}
\end{table*}
All models have relatively stable coefficient estimations showing no evident interaction effects between the regressors (correlation matrix presented in Table~\ref{tab:corrmatrix} in the Appendix). Coefficients should be interpreted relative to each other as opposed to in absolute terms. Because of the relatively small sample size, we refrain from drawing direct conclusions on the model coefficients. For this reason statistical significance is better served in the analysis reported in Figure~\ref{fig:correlation} and is only detailed in Table~\ref{tab:reg} for the reader's reference. Within our sample, model coefficients can be interpreted as the relative change in number of clicks for every additional vulnerability trigger of that type in an email. For example, the M7 coefficient for \texttt{Scarcity} ($0.02$) indicates an increase of $2\%$ in the number of expected clicks for every new trigger of that category. Likewise, an increase in one point on the Levenshtein distance scale is related to a \emph{decrease} in clicks of $10\%$. {A first informal look at the McFadden's $Pseudo-R^2$s, \texttt{Reciprocity}, \texttt{Consistency}, and \texttt{Spoof dist.} appear to have the strongest effect in increasing the explanatory power of the model.}   
\texttt{Scarcity} appears to contribute modestly, whereas \texttt{Liking} appears to have the smallest effect on the model. The negative effect of \texttt{Reciprocity} as shown in Figure~\ref{fig:correlation} is confirmed in the model as well. 

\subsection{Cognitive triaging of phishing success}

We now extend the model evaluation to estimate the amount of clicks generated by other emails for which \ORG\ has detected no click (e.g. because no call-back to \ORG\ resources has
originated from the phishing website, remaining therefore invisible
to \ORG's detection infrastructure, ref. Fig~\ref{fig:process}). Recall however that our model estimates 
are likely subject to overfitting issues due to the inevitably small sample size. This only means that predicted outcomes could be unreliable over arbitrarily diverse email corpora (i.e. not represented in the training data); on the other hand, predictions over \emph{similar} emails to those provided to the fitted models will not suffer from unmodelled biases and will generate reliable estimations. For this reason we only limit our analysis to emails with a distribution of vulnerability triggers within plus or minus one standard deviation from the mean for that trigger in the model's respective training set. 

{To choose the model for the prediction we perform a set of ANOVA tests ($\chi^2$), which indicate all factors add significant information to the model, albeit \texttt{Liking} only marginally. However, due to the statistical limitations of estimations in our dataset, we also consider a second model that considers only isolated factors for which we observe a clear effect as reported in Figure~\ref{fig:correlation}. Based on these observations we consider two different prediction models (PM), each with different regressors, namely: PM1: \texttt{Reciprocity}, \texttt{Consistency}, \texttt{Scarcity} and \texttt{Spoofing distance}; PM2 all six cognitive vulnerabilities + \texttt{Spoofing distance} (i.e. equal to M7 as suggested by the ANOVA tests). This leaves us with $n=334$ and $n=189$ suspicious emails on which to run the predictions for PM1 and PM2 respectively.} To build robust confidence intervals around the estimations, we run a bootstrap simulation ($n=5,000$). Table~\ref{tab:bootres}
\begin{table}
\centering
\footnotesize
\caption{Bootstrapped regression coefficients}
\label{tab:bootres}
\begin{tabular}{lrrrrrr}
  \toprule
 & \multicolumn{3}{c}{\textbf{PM1}} & \multicolumn{3}{c}{\textbf{PM2}}\\
\cmidrule(lr){2-4} \cmidrule(lr){5-7}
 & 0.025q & Med & 0.975q & 0.025q & Med & 0.975q \\ 
  \midrule
$\alpha$ & 3.37 & 4.35 & 4.90 & 2.84 & 4.22 & 5.17 \\
  \texttt{Recip.} & -0.05 & -0.01 & 0.00 & -0.08 & -0.02 & 0.00\\
  \texttt{Cons.} & 0.00 & 0.01 & 0.04 & 0.00 & 0.01 & 0.05 \\
  \texttt{Soc.Pr.} & & & & -0.18 & 0.10 & 0.37\\
  \texttt{Auth}  & & & & -0.03 & 0.00 & 0.05  \\
  \texttt{Scar.} & 0.00 & 0.02 & 0.04 & -0.03 & 0.02 & 0.05 \\
  \texttt{Liking} & & & & -0.02 & 0.04 & 0.05 \\
  \texttt{Sp.dist.} & -0.17 & -0.09 & 0.03 & -0.12 & -0.10 & 0.18\\
  \bottomrule
\end{tabular}
\end{table}
reports median coefficients and 95\% confidence intervals of the estimations. Notice that the estimated coefficients remain largely similar to those of the original models for most coefficients. {PM1 shows much tighter confidence intervals for the estimated coefficients in comparison with PM2, suggesting more reliable predictions. Notice that the distribution of the coefficient estimations in PM1 tends to remain on the same side of zero, again suggesting statistically robust results for this model. This suggests the exclusion of \texttt{Liking}, \texttt{Authority} and \texttt{Social proof} in PM1 may lead to more realistic estimations.}

\begin{table}[t]
\centering
\caption{Descriptive statistics of average predicted clicks}
\label{tab:pred}
\smallskip
\begin{minipage}{0.45\textwidth}
	\footnotesize
	Estimations are generated from 50,000 simulations run on the bootstrapped model coefficients (Table~\ref{tab:bootres}).
	\smallskip
\end{minipage}
\footnotesize
\begin{tabular}{rrrrrr}
\toprule
 \multicolumn{6}{c}{\textbf{PM1}} \\
 
 Min. & 1st Qu. & Median & Mean & 3rd Qu. & Max.   \\ 
\midrule
30 &  48 &  54 &  56 &  62 & 99 \\
\bottomrule\\

\toprule
\multicolumn{6}{c}{\textbf{PM2}}\\

Min. & 1st Qu. & Median & Mean & 3rd Qu. & Max.\\
\midrule
26 &  43 &  50 &  53 &  60 & 129 \\ 
\bottomrule
\end{tabular}
\end{table}

\begin{figure*}[t]
\centering
\includegraphics[width=0.80\textwidth]{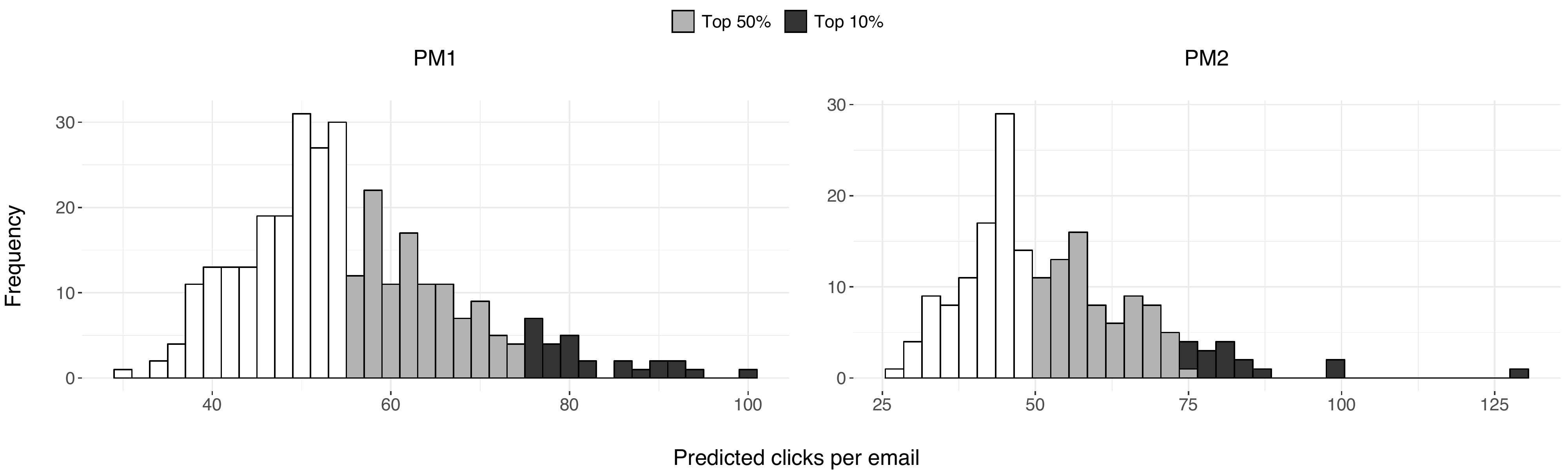}
\caption{Distribution of predicted average clicks}
\label{fig:predclicks}
\end{figure*}
 We simulate model predictions for the undetected clicks by
randomly sampling ($n=50,000$) model coefficients from the two distributions and report aggregate statistics (Table \ref{tab:pred}) of the estimated number of generated clicks. Figure \ref{fig:predclicks} reports the results.
The simulation results indicate that the average `undetected' email has potentially generated 50 - 55 clicks, with a long tail of (few) emails generating up to 100 clicks.\footnote{Notice that additional organization-specific features of the email (e.g. presence of the company logo), may also have an effect on the number of clicks. Whereas this is out of the scope of this paper, which only looks at the cognitive effects, a fully-operative model within an organization can easily integrate other factors in the prediction.} This suggests that prioritization efforts based on the cognitive characteristics of a phishing email could help in more efficiently addressing
attacks (e.g. by means of takedown actions), by targeting first attacker resources that are likely to generate more impact on the organization's customer base: by targeting first the emails that are most likely to engage users in compliant behaviour, organizations can effectively triage the stream of incoming phishing attacks to minimize the impact on their customer base.

\section{Discussion}
\label{sec:discussion}
The previous sections have demonstrated how quantitative measurements of cognitive vulnerabilities employed in phishing attacks can be used to develop a model to make predictions about the expected efficacy of these attacks. This characterization allows one to assess the threat of these attacks in an automated way such that instant prioritization of phishing incident responses becomes possible. This paper's contributions go beyond the scope of earlier works on cognitive factors for phishing by providing an empirical estimation and operable implementation to estimate phishing success.

In this work we identified several correlations between different cognitive vulnerabilities and the average number of clicks an email can be expected to generate. {In line with the hypothesis that the presence of any individual cognitive vulnerability increases user response to the phish, we found that \texttt{Consistency} and \texttt{Scarcity} exercise a clear positive effect on the number of generated clicks. We find no evident effect from \texttt{Social proof} and \texttt{Liking}, 
whereas  \texttt{Authority} appears to have a positive effect albeit driven by only a few non-zero data points. Interestingly, \texttt{Reciprocity} even shows a counterproductive effect, albeit only marginal.} 
This difference may well be explained by the specific application domain, as corporate customers subject to financial threats from phishing can generally be expected to have different sensitivity to specific {principles of influence} than other groups~\cite{Lawson2017}. Although this suggests that full generalizability can not be expected for any one set of results, conclusions similar to ours could be drawn for specific contexts close in nature to the one in which \ORG\ operates. {In particular, our finding of the reduced effect of \texttt{Authority} in the banking domain is in contrast with results from Wash and Cooper~\cite{Wash2018}, who found \texttt{Authority} to be the most effective strategy for the presentation of certain phishing education materials. This difference may illustrate the context-dependency of the relative efficacy of these influence tactics, and indicates a need for careful consideration of such differences across domains. 
For example, the effect of \texttt{Authority} can be mediated here by the already relatively high authoritive position a bank has on its customers; this suggests that, on one side, depending on the domain it may be more difficult for an attacker to devise effective attacks \emph{adding} to baseline cognitive effects; on the other, this also suggest that a relevant metric to evaluate could be the \emph{relative increase} (or decrease) in the cognitive effect w.r.t. the baseline. A similar consideration could be drawn for \texttt{Reciprocity}, whereby we observe a negative effect on the generated `clicks'. An explanation could be that these type of triggers rise a red flag in the context of banking operations, for example as a bank's `environmental friendliness' may not be a convincing-enough reason to act on a request (e.g. to renew one's debit card).} 

These observations also provide useful input to training campaigns regularly run by medium and large organizations in an attempt to increase their customers and employee's awareness of the social engineering threat. Replications of this study in specific domains could reveal which principles of influence the `average' customer
of an organization is more vulnerable to; awareness campaigns run by the organization could then target those specific traits by providing specific examples or information material built ad-hoc for the consumer base (or targeting sections of it). For example, consumers particularly vulnerable to \texttt{Scarcity} may benefit from knowing the organization's policies in terms of change deadlines and processes, such that an email stating unrealistic and short cutoff dates to react lose credibility. 

Operationally, the presented procedure could be applied both client and server side to automate the risk evaluation of potential phishing emails for the enforcement of security policies; {for example, mail client plugins or server-side processes could automatically divert or forward high-risk emails to phishing investigation and response teams for further evaluation, while delaying the delivery of messages waiting for a diagnosis}.
Furthermore, we have described how these observed effects can be used in the construction of a prediction model for the triaging of incoming phishing attacks. 
By enabling the triaging of incoming phishing attacks, our results will enable incident response teams to focus on the most prominent threats immediately, without having to manually filter out the noise from the bulk of low priority emails in their phishing abuse inbox, thereby minimizing reaction costs and increasing response effectiveness. The practicality of this is evidenced in Figure~\ref{fig:predclicks}, where by addressing the small fraction of emails associated with the highest expected click counts one can mitigate a large fraction of potential attacks. {This is critical to minimize overall victimization rates, as the short-lived nature of phishing domains stresses a need for prompt identification of which domains are most likely to be reached by customers falling for the phish. By contrast, addressing attacks in no particular order would most likely result in wasting valuable time and resources by addressing first the vast majority of attacks that are likely to generate few clicks only (ref. Figure~\ref{fig:predclicks} and Table~\ref{tab:pred}).} 

Finally, our method opens up new opportunities in terms of automated incident handling and security orchestration, e.g. by enabling incident handlers to apply automated follow up procedures to incoming phishing attacks that fall within a certain threat range; for example, reported measures on the vulnerability triggers could be used to implement dynamic risk-based access control policies to limit immediate follow-up actions. Similarly, CSIRTs (\textit{Computer Security Incident Response Team}) could implement automated network-level containment procedures based on the profile of incoming emails, and avoid additional (and unnecessary) victimization by delaying follow-up actions by the users until the risk is cleared.

\section{Conclusions}
\label{sec:conclusions}

In this work we presented an empirical method and evaluation of the effect of cognitive vulnerability triggers in phishing emails on the expected `success' of an attack. We employed a unique dataset from a large European financial organization with data from
their phishing response division. Our results indicate that response teams' operations, such as take-down actions against rogue phishing domains, could largely benefit from a (fully automated) cognitive assessment of the email body to predict relative success
of the attack, given the relevant user base. Our findings and method could also be employed to deploy more effective training and awareness campaigns in response to the more prominent threats suffered by the potential victims. Future work could explore automated response strategies to contain potential attacks and/or delay user response where needed.


{\normalsize \bibliographystyle{plain}
\bibliography{library.bib}}

\appendix

\section*{Appendix}

\begin{table}[h!]
\centering
\scriptsize
\caption{Correlations between regression variables} 
\label{tab:corrmatrix}
\begin{minipage}{0.95\columnwidth}
	\footnotesize
	We find no evident correlations between regressors. Only \texttt{Scarcity} and \texttt{Social proof}, and \texttt{Liking} and \texttt{Spoof dist.} show a higher than average correlation of 0.50 and 0.57 respectively, which is unlikely to affect estimation results.
	\smallskip
\end{minipage}
\begin{tabular}{lrrrrrrr}
  \toprule
 & (1) & (2) & (3) & (4) & (5) & (6) & (7) \\ 
  \cmidrule{2-8}
  (1) \texttt{Reciprocity} & 1.00 & -0.19 & 0.13 & -0.06 & -0.11 & -0.09 & -0.17 \\ 
  (2) \texttt{Consistency} &  & 1.00 & -0.08 & -0.24 & 0.10 & -0.09 & -0.41 \\ 
  (3) \texttt{Social proof} &  &  & 1.00 & 0.25 & -0.04 & 0.50 & 0.13 \\ 
  (4) \texttt{Authority} & &  &  & 1.00 & 0.06 & -0.08 & -0.10 \\ 
  (5) \texttt{Liking} &  & &  &  & 1.00 & 0.09 & 0.57 \\ 
  (6) \texttt{Scarcity} & &  &  &  &  & 1.00 & 0.24 \\ 
  (7) \texttt{Spoof dist.} &  &  &  &  &  &  & 1.00 \\ 
   \bottomrule
\end{tabular}
\end{table}

\subsection*{Bootstrap analysis for similar email detection}

\begin{figure}[H]
  \centering
  \includegraphics[width=0.70\columnwidth]{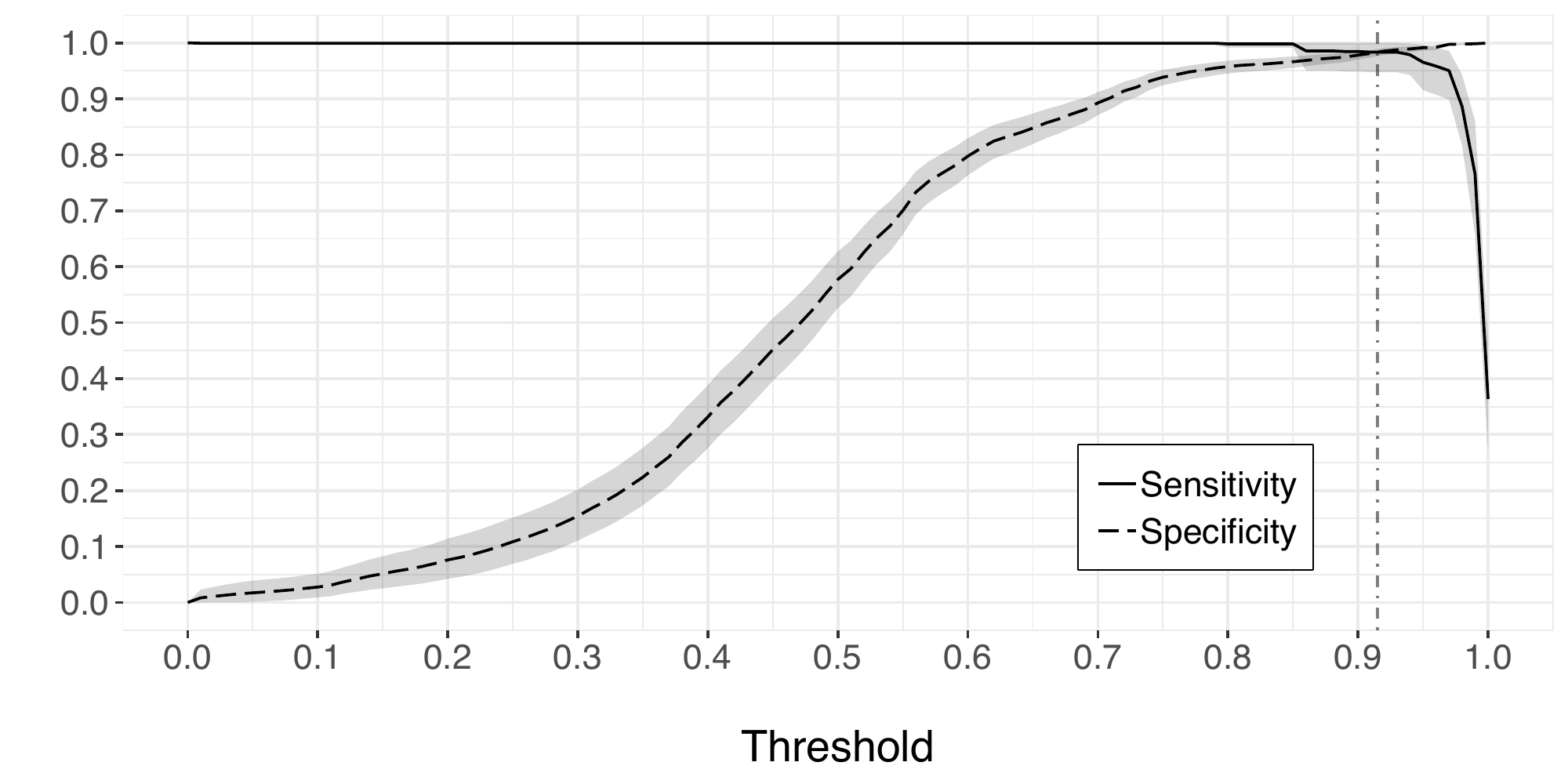}
  \begin{minipage}{0.45\textwidth}
	\smallskip
  	\footnotesize
  	The optimal threshold was found at 0.91 based on the intersection of the mean sensitivity and specificity metrics at all decimal thresholds in [0,1] across 10,000 bootstrap simulations with sample size 300.
  \end{minipage}
  \caption{Simulated optimal cosine similarity threshold for duplicate detection}
  \label{figure:threshold}
\end{figure}

After computation of the full pairwise similarity matrix for all suspect emails in our dataset, a threshold value was used to determine the lower-bound for the similarity score of emails we consider to be duplicates. In order to determine the most optimal threshold value for our specific dataset we performed a \textit{bootstrap analysis} \cite{Efron1993}. A bootstrap analysis generally involves repeatedly running simulations on samples drawn with replacement from an original sample set in order to estimate statistics on a larger population. This is a fitting solution for problems concerning dataset of large sizes like ours, which do not generally allow for efficient derivation of the full set of results that qualify as \textit{``ground-truth''}. 

We started our bootstrap analysis with a random sample of 300 suspect phishing emails for which we made a manual assessment of all pairwise similarities to test the performance of our cosine similarity algorithm across different thresholds. Then, we repeatedly ($n = 10,000$) drew samples with replacement of size 300 from our manually classified sample and computed the pairwise cosine similarity matrix for all decimal thresholds in the interval [0,1]. For each combination of bootstrap sample and threshold value we computed the performance using the sensitivity (true positive rate) and specificity metrics (true negative rate). 
A high sensitivity score refers to a high probability of duplicate detection, measured by the proportion of actual duplicates that are correctly identified as being similar, whereas a high specificity score refers to a high probability of non-duplicate rejection, measured by the proportion of actual non-duplicates that are correctly identified as not being similar. The intersections of the mean results for these two performance measures indicate that 0.91 is the optimal threshold value for our dataset, as is visualized more elaborately in Figure \ref{figure:threshold}.

We use this threshold to calculate the pairwise similarity matrix for all emails in our dataset and assign emails that are found to be similar the same \textit{duplicate ID} to allow us to filter for unique emails.



\subsection*{Distribution of phishing emails by user}

\begin{figure*}[t]
\centering
\includegraphics[width=0.48\textwidth]{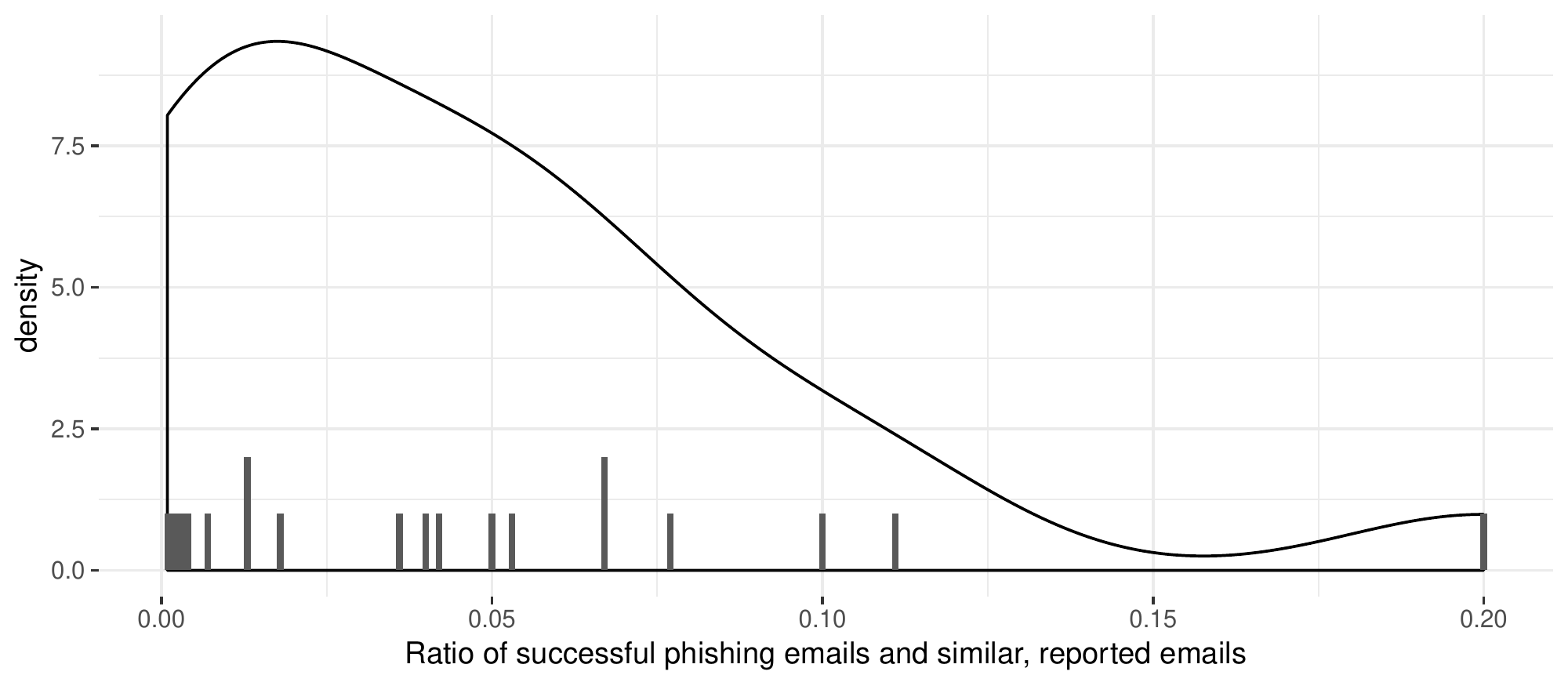}
\includegraphics[width=0.48\textwidth]{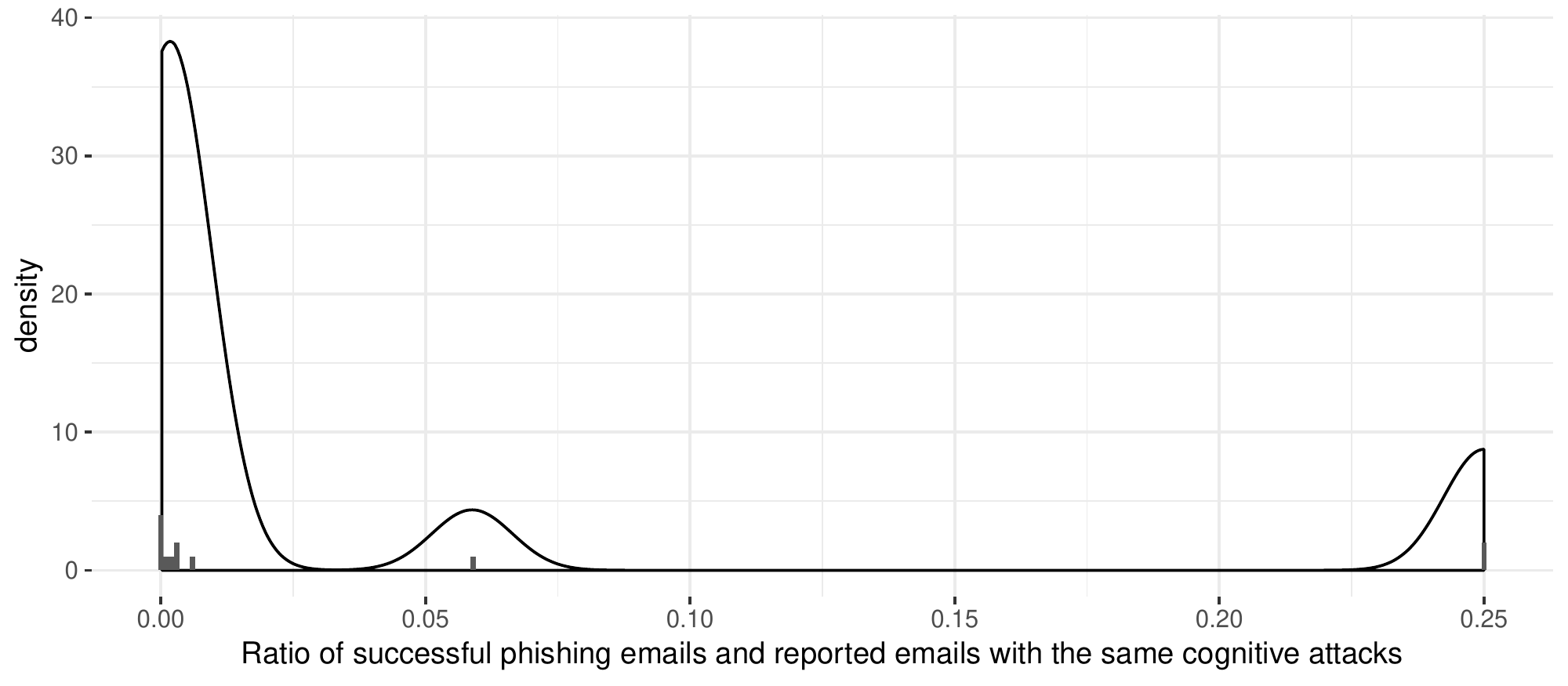}
\begin{minipage}{0.85\textwidth}
\footnotesize
The distribution is calculated over similar emails (as per their cosine similarity, left), and over emails that share the same cognitive attacks (right). We observe a stable ratio across all emails, suggesting no significant skew in the probability of an email reaching a user's inbox.
\end{minipage}
\caption{Estimation of rates of arrival of phishing emails to users}
\label{fig:estiamtes}
\end{figure*}
\begin{figure}[t]
\includegraphics[width=0.9\columnwidth]{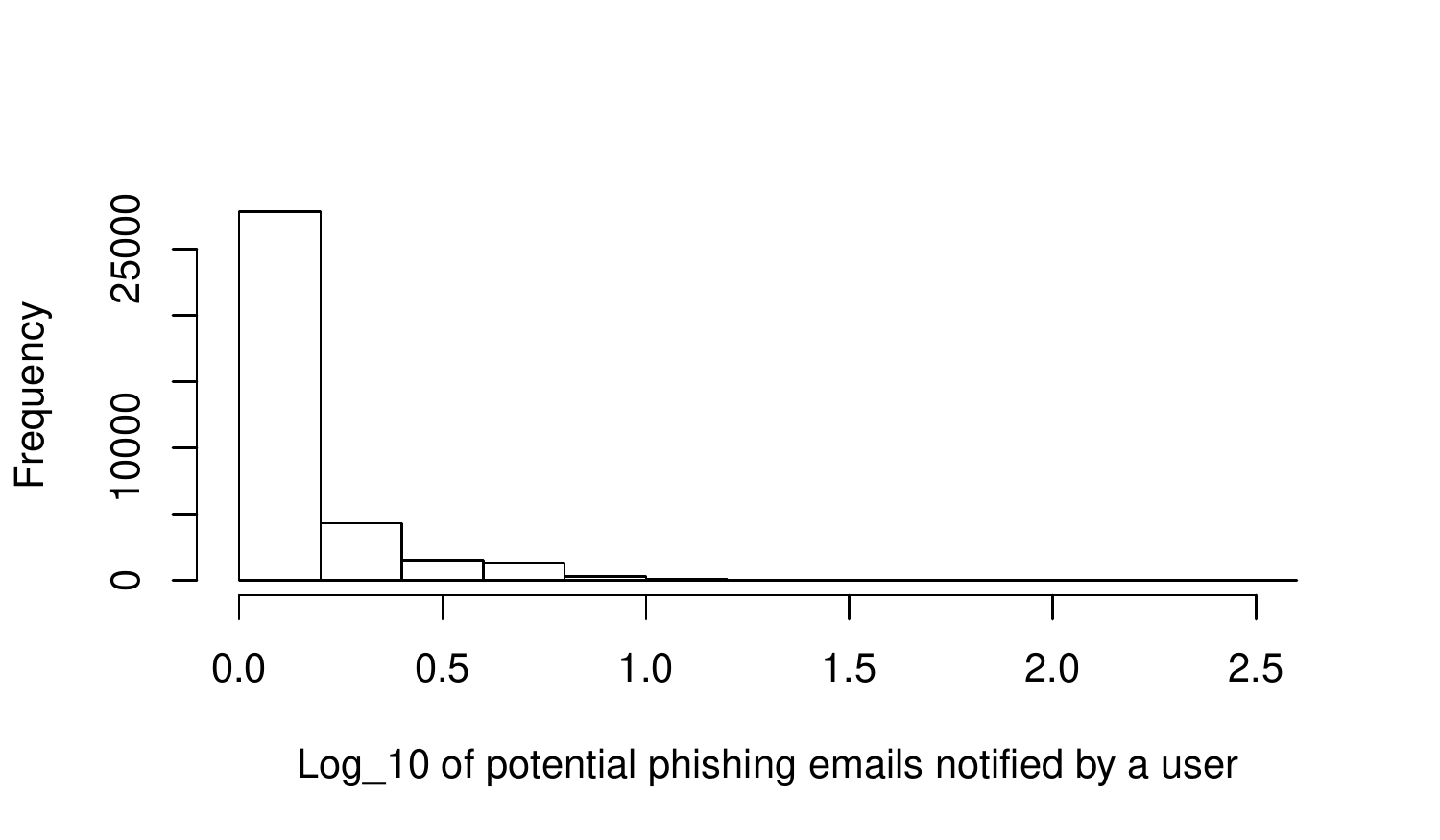}
\caption{Number of users reporting $|D_u|$ emails ($log_{10}$)}
\label{fig:notifications}
\end{figure}

{We perform a robustness check to evaluate the robustness of our results against large biases in the distribution of phishing emails across potential victims (whereby higher reported clicks may relate to higher email delivery volumes). We base the following on the sole assumption that attackers “sample” victims from the same pool (i.e. \ORG\ customers). As we of course cannot measure email delivery rates in user inboxes, our aim is to ask how would the dataset look like if large sample biases were present, and look for evidence in the data. 

We developed the following data generation model to formalize and test this:
a phishing email $e \in E$ can reach a user $u \in U$ with probability $P_e(u)$. 
A suspicious email will be detected by a user with a certain probability $P^{det}_{u,e}=P_u(DET|e)$. Notice that this depends on the email and the specific user that receives it, as different users may have different sensibilities to emails with different characteristics. The probability of remaining undetected is simply the complement, and is defined as $P^{und}_{u,e}=1-P_u(DET|e)$. 

Further, each user has a certain probability $P_u(notify|DET,e)$ and $P_u(click|\neg DET, e)$ of, respectively, notifying a detected email, and clicking on a link if the email is not detected.
Hence, the probability of an email being reported by a user is $P_e(u)\cdot P_u(DET|e) \cdot P_u(notify|DET, e)$. Conversely, $P_e(u)\cdot (1-P_u(DET|e))\cdot P_u(click|\neg DET, e)$ is the probability of a click for each $e \in E$ and $u\in U$. 

Let $C$ be the set of clicked emails, and $D$ the set of reported emails, we'd then have:

\begin{eqnarray}
\label{eq2}|C| &=& |E|  \sum_{\forall e \in E} \sum_{\forall u \in U} P_e(u) \cdot P^{und}_{u,e}\cdot  P_u(click|\neg DET, e)\\
\label{eq1}|D| &=& |E|  \sum_{\forall e \in E} \sum_{\forall u \in U} P_e(u) \cdot P^{det}_{u,e} \cdot P_u(notify|DET, e)
\end{eqnarray}

$|D|$ corresponds to the whole set of suspicious emails reported in the organization's phishing inbox. $|C|$ corresponds to the set of emails clicked (of which we observe $C' = D \cap C$).




Notice that $P_e(u)$ (i.e. the probability of an email arriving to a user's inbox) is the only variable that is outside of the direct influence of the user. On the contrary, $P^{det}_{u,e}$, $P^{und}_{u,e}$, $P_u(notify|DET, e)$ and $P_u(click|\neg DET, e)$ directly depend on the {characteristics of the user and of the email}. Hence, we would expect them to be approximately constant as long as within the comparison the users are the same, and the emails are similar to each other. As we can control for `similar' emails (Section~\ref{sec:dupldetect}) and users are all pooled from the set of \ORG's clients, we can  isolate large effects on $|C'|$ and $|D|$ as being caused by large fluctuations in $P_e(u)$. 

Specifically, we would expect $|C'_{similar}|/|D_{similar}| \approx constant$ \textbf{iff} also $P_e(u)\approx constant\ \forall e \in E_{similar}$ as all other terms in the two equations will remain approximately the same for similar emails and the users  sampled from the same pool.
Hence, we measure the ratio of $|C'_{similar}|/|D_{similar}|$ as a {proxy} to estimate how much $P_e(u)$ can be expected to vary across emails. This holds under the sole assumption that emails in $D \setminus C$ are indistinguishable (from the perspective of the user) from those in $D\cap C=C'$; this is uncontroversial as the detection mechanism for the inclusion of an email in $C'$ only depends on the phishing webpage and not on the email per se.

Figure~\ref{fig:estiamtes}
reports the ratio distribution calculated over emails with cosine similarity above the defined threshold (left), and over emails employing the same cognitive attacks (right). The ratios are small (i.e. only a small fraction of reported emails of a certain type can be expected to generate at least one click). This is qualitatively and quantitatively in line with previous findings in the literature~\cite{Kanich-2008-CCS}. Importantly, we observe that under both measures of similarity the rate is essentially constant and settles around 1-5\% for all emails. This observation is incompatible with a significantly skewed distribution of emails per user. Breaking down the figure by users receiving the phishing email does not reveal any additional pattern as most users report only a few emails each (ref Figure~\ref{fig:cdftos} and Figure~\ref{fig:notifications}).

This is in line with previous literature on phishing attacks \cite{Dhamija2006, ho2017detecting} suggesting that no specific pre-selection of users charcterizes untargeted phishing attacks. We therefore do not expect significant biases in the analysis to emerge by the otherwise unmeasurable distribution of emails in users' inboxes.}





\end{document}